\documentclass[aps,prb,twocolumn,showpacs,amsmath,amssymb,superscriptaddress]{revtex4-2}
\usepackage{graphicx}
\usepackage{CJK}
\usepackage{color}
\usepackage[colorlinks,bookmarks=false,citecolor=blue,linkcolor=red,urlcolor=blue]{hyperref}
\usepackage{multirow}
\usepackage{ulem}
\usepackage{tabularx}
\usepackage[nounderscore]{syntax}
\graphicspath{{figs/}}
\usepackage{subfigure}
\usepackage{float}

\newcommand{\be}{\begin{equation}}
\newcommand{\ee}{\end{equation}}

\begin{document}
\begin{CJK*}{UTF8}{gbsn}
\title{Boundary sensitive Lindbladians  and relaxation dynamics}

\author{Xu Feng}
\affiliation{Beijing National Laboratory for Condensed Matter Physics, Institute
of Physics, Chinese Academy of Sciences, Beijing 100190, China}
\affiliation{School of Physical Sciences, University of Chinese Academy of Sciences,
Beijing 100049, China }

\author{Shu Chen}
\email{schen@iphy.ac.cn }
\affiliation{Beijing National Laboratory for Condensed Matter Physics, Institute
of Physics, Chinese Academy of Sciences, Beijing 100190, China}
\affiliation{School of Physical Sciences, University of Chinese Academy of Sciences,
Beijing 100049, China }

\date{\today}
\begin{abstract}

It is well known that non-Hermitian systems  can be extremely sensitive to boundary conditions owing to non-Hermitian skin effect (NHSE).  Analogously, we investigate two  boundary-sensitive $U(1)$ symmetric  Lindbladians: one carries current in the steady state, and the other does not. 
The numerical results indicate significant change of the Liouvillian spectrum,  eigenmodes and relaxation time for both Lindbladians when the boundary conditions are altered. This phenomenon  is found to be triggered by the Liouvillian skin effect (LSE), specifically the localization of eigenmodes, which stems from the NHSE of the non-Hermitian effective Hamiltonian. In addition, 
 these two Lindbladians manifest different  LSE, ultimately resulting in distinct relaxation behaviors.

\end{abstract}
\maketitle


\section{Introduction}
\label{intro}

Open quantum systems have attracted growing interest due to the rapid advancements in quantum technology, which enable the engineering of specific environments and the realization of diverse non-equilibrium phenomena \cite{OQS,QuantumControl,TrappedIons,AlgebraicExponentialDecoherence,diehlQuantumStatesPhases2008,diehlTopologyDissipationAtomic2011,siebererKeldyshFieldTheory2016,QuantumComputationQuantumstate2009,RydbergQuantumSimulator2010}. Moreover, given the inevitable coupling between the system and the environment in practical applications, it is both essential and practical to explore physical processes encompassing system-environment interactions.  Under Markovian approximation, it is widely acknowledged  that the Lindblad master equation effectively describes the evolution of the system\cite{lindblad}. One  pivotal quantity that characterizes this evolution is  the relaxation time $\tau$, serving as a significant intrinsic timescale.  Another closely related quantity is   Liouvillian gap $\Delta$, which is defined as the smallest modulus of the real part of nonzero eigenvalues of the Lindbladian.
Conventionally, for open
quantum systems  governed by the Markovian Lindblad master equation, the relaxation time $\tau$ is approximately 
inversely proportional to the Liouvillian gap $\Delta$ of the
system. Therefore, extensive prior research has focused on investigating the scaling relationship between the Liouvillian gap and system size \cite{vernierMixingTimesCutoffs2020,zhouExponentialSizeScaling2022,znidaricRelaxationTimesDissipative2015}. However, recent studies have uncovered the discrepancy between the inverse of the Liouvillian gap and the relaxation time \cite{moriResolvingDiscrepancyLiouvillian2020,hagaLiouvillianSkinEffect2021,wangAcceleratingRelaxationDynamics2023,SymmetrizedLiouvillianGap}. 
This anomalous relaxation behavior is attributed to  superexponentially
large expansion coefficients of  Liouvillian eigenmodes.

On another front, recent works have uncovered numerous novel phenomena of non-Hermitian Hamiltonian, one of which is the extreme sensitivity of the boundary conditions, which originates from NHSE \cite{GBZ, photonic, WindingSkin,windingSkin2, Anatomyskinmodes,boundarymodes,non-bloch,auxiliaryGBZ,boundarysensitive, CNHSE,zhangUniversalNonHermitianSkin2022,wangAmoebaFormulationNonHermitian2022,huNonHermitianBandTheory2023,manybodyNHSE,symtopNH,ashida2020non,exceptional,linTopologicalNonHermitianSkin2023,okumaNonHermitianTopologicalPhenomena2023}. There are two marked signals for NHSE: one is accumulation of extensive eigenstates near the edge, and another is dramatic change of energy spectrum when switching boundary conditions. 
Evidently, NHSE will induce rich nontrivial dynamical effects, such as unconventional reflection, entanglement suppression, etc \cite{songNonHermitianSkinEffect2019,HelicalDamping,yangLiouvillianSkinEffect2022,liDynamicSkinEffects2022,NHSEEE,sawadaRoleTopologyRelaxation2023,mcdonaldNonequilibriumStationaryStates2022,liNonBlochDynamicsTopology2023}. However, the non-Hermitian Hamiltonian suffers from  post-selection problem \cite{PhysRevResearch.4.L032026}, which renders it exponentially hard to implement in experiments.  
So it is natural to ask how to design a post-selection free open quantum model hosting skin effect and how skin effect will shape the Liouvillian spectrum, eigenmodes (including steady state) and the relaxation dynamics?

The first question is answered  inspired by recent progress on measurement-induced phase transition (MIPT) \cite{Nahum, FisherPRB1, FishePRB2, unitaryproj,liuUniversalKPZScaling2022,fermionMIPT, DiehlMIPT,Isingchain,purificationMIPT}, which similarly encounters the post-selection issue.  An interesting attempt to overcome the problem is to apply corrective unitary operations, as long as the unexpected  measurement outcome appears. In doing so, the system can be steered towards a particular target state \cite{buchholdRevealingMeasurementinducedPhase2022,odeaEntanglementAbsorbingStateTransitions2022,ravindranathEntanglementSteeringAdaptive2022,ravindranathFreeFermionsAdaptive,sierantControllingEntanglementAbsorbing2022}. By choosing appropriate continuous measurement and feedback operators, recent studies  successfully attain Liouvillian skin effect (LSE) in the steady state. The numerical findings corroborate that the skin effect effectively immobilizes a substantial portion of particles, thereby significantly suppressing the growth of entanglement and ultimately preventing the occurrence of MIPT\cite{wangAbsenceEntanglementTransition2023,fengAbsenceLogarithmicAlgebraic2023}. It is worthwhile to mention other mechanisms such as interactions \cite{hamanakaInteractioninducedLiouvillianSkin2023} and asymmetric jump operators can also induce LSE\cite{hagaLiouvillianSkinEffect2021,ManyBodyNHSEgaugecoupling2023}. 

In this paper, we will address the second question.  In order to clearly show the rich and nontrivial role of LSE, we investigate two Lindbladians, one with feedback and the other without,  under both periodic boundary conditions (PBCs) and open boundary conditions (OBCs). The first Lindbladian holds a  current-carrying steady state, whereas the second Lindbladian hosts a no-current steady state. It is  reasonable to anticipate  these two Lindbladian will manifest distinct relaxation behaviors.
Our paper indeed demonstrates that LSE will greatly alters the Lindbladians' eigenvalues and eigenmodes. Consequently, the relaxation behaviors are also fundamentally changed.  Furthermore, the Lindbladians studied in this paper can be well understood by perturbation theory, which unveils the close connections between the NHSE of non-Hermitian effective Hamiltonian and the LSE of the whole Lindbladian.  For clarity, we adopt NHSE to  denote  skin effect for the eigenstates of the non-Hermitian Hamiltonian, while LSE represents skin effect for the eigenmodes of the Lindbladians, despite both of them sharing the same mathematical origin.

The rest of the paper is organized as follows: in Sec.  \ref{model}, we introduce  two Lindbladians  and their corresponding measurement protocols. Next, Sec. \ref{KnowledgeLindbladian} provides a brief overview of some basic knowledge of Lindbladians.
Subsequently, we proceed to present our research findings.
 Firstly, in Sec. \ref{withfeedbcak}, the Lindbladian with feedback is investigated. Specifically, in Sec. \ref{feedbackspectrum}, we  acquire Liouvillian spectra  for different monitoring rate $\gamma$ under OBCs and PBCs. Secondly, the Sec. \ref{liouvillianskineffect} shows that the eigenmodes including steady state under OBCs exhibit LSE, while the eigenmodes under  PBCs are extended.   Thirdly, in Sec. \ref{relaxationdynamics}, utilizing exact diagonalization, we obtain scaling relation between the  Liouvillian gap $\Delta$ and system size $L$ for OBCs and PBCs. We find the relaxation time scales as $\tau \sim O(L)$ under OBCs, and $\tau \sim O(L^2)$ under PBCs. The former displays discrepancy between the inverse of the Liouvillian gap
and the relaxation time, which  approximately follows  $\tau\sim 1/\Delta+L/\xi\Delta$, while the latter  obeys $\tau\sim 1/\Delta$.  Interestingly, cutoff phenomenon exists under OBCs. In Sec. \ref{feedbackmanybody}, we consider the half-filled many-body case in brief, in which the LSE and cutoff phenomenon survive. The scaling relations between $\tau$ and $L$ remain valid. In Sec. \ref{nofeedback},   the Lindbladian without feedback is investigated. Although the steady state and eigenmodes around steady state are free from LSE, the majority of eigenmodes still display LSE. Hence, the Liouvillian spectrum, Liouvillian gap and relaxation time are also dramatically altered when changing boundary conditions. Interestingly, we observe that the relaxation time $\tau$ is smaller than the inverse of Liouvillian gap under PBCs, i.e. $\tau < 1/\Delta$, while for the OBCs, the relaxation time still satisfies the usual rule $\tau\sim 1/\Delta$.
Appendix. \ref{sec:appendixA}  presents a perturbation analysis of Liouvillian spectrum and eigenmodes. Exact solutions of the nontrivial steady state under PBCs are also  available. Lastly, in Appendix. \ref{additionalNumericalResults}, some additional numerical results are provided to support our  conclusions. The main results of the paper are summarized in table. \ref{summary}.

\begin{table*}[bt]
\begin{center}
    \centering
    \begin{tabular}{p{30mm}p{41mm}p{21mm}p{17mm}p{18mm}p{18mm}p{25mm}}\hline\hline
        \ & steady & the majority  & relaxation  & relaxation & cutoff & Liouvillian gap vs \\
         \ &state & of eigenmodes & time vs $L$ & time vs $\gamma$ & phenomena &  relaxation time \\  \hline
        Feedback, PBCs & $\sum_{k}\lambda_{k}|k\rangle\otimes|k\rangle^{*}$& extended & $O(L^2)$& monotonic  & not exist & $\tau \sim \dfrac{1}{\Delta}$ \\
        Feedback, OBCs & LSE & LSE & $O(L)$ & non-monotonic & exist & $\tau\sim \dfrac{1}{\Delta}+\dfrac{L}{\xi\Delta}$ \\
        No-feedback, PBCs &  $\Bbb{I}/L$ for $L\neq 4N$; \  $\Bbb{I}/L,  |-\dfrac{\pi}{2}\rangle\otimes|-\dfrac{\pi}{2}\rangle^{*}$ for $L=4N$ & extended & $O(L^\alpha)$,\ \ \ \ \ \ \ $\alpha <2$ & monotonic & not exist & $\tau < \dfrac{1}{\Delta}$  
        \\ 
        No-feedback, OBCs & $\Bbb{I}/L$ & LSE & $O(L^{\beta})$ &  non-monotonic & not exist & $\tau\sim\dfrac{1}{\Delta}$  \\
        \hline\hline
    \end{tabular}
    \caption{
    Summary of the main results of the paper. ``LSE", abbreviation for Liouvillian skin effect,  represents the skin localization of the eigenmodes and $\xi$ characterizes the localization length.  The steady state for the Lindbladian with feedback under PBCs is $\rho_{ss}=\sum_{k}\lambda_{k}|k\rangle\otimes|k\rangle^{*}$, in which $\lambda_{k}=\dfrac{1-\text{sin}k}{1+\text{sin}k}/{\sum_{k}\dfrac{1-\text{sin}k}{1+\text{sin}k}}$, $k=\dfrac{2\pi j}{L}, j=1,2...L$, and  $|k\rangle=\dfrac{1}{\sqrt{L}}\sum_{j=1}^{L}e^{-ikj}|j\rangle$. For the Lindbladian without feedback under PBCs, if system size $L=4N, N\in\mathcal{Z}$, there exists bistable steady states $\Bbb{I}/L$ and $|-\dfrac{\pi}{2}\rangle\otimes|-\dfrac{\pi}{2}\rangle^{*}$. Otherwise, the steady state is the unique maximally mixed state $\Bbb{I}/L$. Moreover, the relaxation time approximately obeys $\tau\propto O(L^{\alpha})$, in which $\alpha$ is dependent on $\gamma$ and smaller than 2 due to faster relaxation process. For no-feedback case under OBCs, the relaxation time follows $O(L^{\beta})$, in which $\beta$ is given by the scaling behavior of Liouvillian gap  $\Delta\propto L^{-\beta}$ and $\beta$ depends on $\gamma$.
    The initial states are all chosen as  $|L\rangle\langle L|$ for the above four cases.
    }
    \label{summary}
    \end{center}
\end{table*}

\section{Lindbladians and Measurement Protocols}
\label{model}
The measurement protocols to achieve the Lindbladians we studied have been illustrated in detail before \cite{wangAbsenceEntanglementTransition2023,fengAbsenceLogarithmicAlgebraic2023}. Here, we will briefly introduce them   for completeness.
We consider a one-dimensional (1D) free fermion model, subject to continuous measurement and  unitary feedback. The Hamiltonian is  given by
\begin{equation}\label{eq1}
\begin{split}
\hat{H}=\sum_{i}\dfrac{t}{4}(c_{i+1}^{\dagger}c_{i}+c_{i}^{\dagger}c_{i+1}),
\end{split}
\end{equation}
where $c_{i}$ ($c_{i}^{\dagger}$) denotes annihilation  (creation) operator of the spinless fermion at site $i$ and  $t$ represents the hopping strength. We set $t=1$ throughout the work. The measurement operators act on two neighboring sites and can be expressed as $\hat{P}_{i}=\xi^{\dagger}_{i}\xi_{i}$, in which  $\xi^{\dagger}_{i}=\dfrac{1}{\sqrt{2}}(c^{\dagger}_{i}-ic^{\dagger}_{i+1})$. To achieve nontrivial steady state exhibiting LSE, additional subsequent unitary feedback operator should be applied after measurement. The unitary feedback operator is expressed as $U_{i}=e^{i\pi n_{i+1}}$. Thus, for the Lindbladian  with feedback, the corresponding Lindblad operator is \begin{equation}
	\begin{split}
	L_{i} &= U_{i}P_{i}= e^{i\pi n_{i+1}}\,  \xi^{\dagger}_{i}\, \xi_{i}.
		\end{split}
	\end{equation}
Furthermore, we also consider the Lindbladian without feedback. 
In this scenario, the Lindblad operator simplifies to the measurement operator alone, represented as  $L_{i}=P_{i}$. It is noted that  $L_{i}^{\dagger}L_{i}$ is invariant with unitary feedback. In addition, both Lindbladians are   number-conserving, or $U(1)$ symmetric.

For continuously monitored systems, it is widely acknowledged that the system's evolution is described by the Lindblad master equation within the framework of the Markovian approximation\cite{fermionMIPT}. The master equation is as follows:
\begin{equation}\label{LME}
\begin{aligned}
	\begin{split}
	\dfrac{d\rho}{dt}&=\mathcal{L}\rho \\
	&=-i[{H}, \rho]+\gamma\sum_{\mu}\left({L}_{\mu}\rho{L}_{\mu}^{\dagger}-\dfrac{1}{2}\{{L}_{\mu}^{\dagger}{L}_{\mu},{\rho}\} \right) \\ 
	&=-i({H}_{\text{eff}}\rho-\rho {H}_{\text{eff}}^{\dagger})+\gamma\sum_{\mu}{L}_{\mu}\rho{L}_{\mu}^{\dagger}
	,
		\end{split}
		\end{aligned}
	\end{equation}
 where $\mathcal{L}$ is referred to as  Lindbladian or Liouvillian superoperator and ${L}_{\mu}$ denotes  Lindblad operator mentioned previously. Furthermore, $\gamma$ represents the monitoring rate, namely the frequency for the environment observing the system. It should be noted that the Lindblad operators  are  quadratic in the single-particle sector. 
 Moreover, we highlight that both  Lindbladians (with and without feedback) share the same effective non-Hermitian Hamiltonian $H_{\text{eff}}$, which can be expressed as 
\begin{equation}
\begin{aligned}
{H}_{\text{eff}}&={H}-i\dfrac{\gamma}{2}\sum_{\mu}{L}_{\mu}^{\dagger}{L}_{\mu}\\ 
&= \dfrac{1}{4}\sum_{i}\left[
 (t+\gamma) {c}^{\dagger}_{i}{c}_{i+1}+(t-\gamma){c}^{\dagger}_{i+1}{c}_{i}- i\gamma({n}_{i}+{n}_{i+1})\right].
\end{aligned}
\end{equation}
Neglecting overall dissipation, $\hat{H}_{\text{eff}}$  corresponds to the Hatano-Nelson model\cite{HNmodel}, renowned for its manifestation of NHSE under OBCs. The eigenstates display exponential localization  characterized by  $r^{-i}$, in which $r=\sqrt{|t+\gamma|/|t-\gamma|}$ and $i=1, 2, ...L$. 
Conversely, under PBCs, owing to $[H, \sum_{\mu=1}^{L}L^{\dagger}_{\mu}L_{\mu}]=0$, the non-Hermitian effective Hamiltonian $H_{\text{eff}}$ shares common  eigenstates $|k\rangle$  with free fermion Hamiltonian $H$. As a result, the eigenvalues are $H_{\text{eff}}|k\rangle=E(k)|k\rangle$, in which $E(k)=\dfrac{1}{2}(\text{cos}k-i\gamma\text{sin}k)-i\dfrac{\gamma}{2}$, and $k=\dfrac{2\pi j}{L}, j=1,2,..L$.
In the Appendix. \ref{sec:appendixA}, perturbation theory explicitly demonstrates that the sensitivity of the Lindbladian  is inherited from the sensitivity of non-Hermitian effective Hamiltonian $H_{\text{eff}}$. 

In this paper, we investigate two Lindbladians, corresponding to feedback case and no-feedback case,  respectively.  Both of them are boundary-sensitive, but  it is noteworthy that  there is an essential difference between the cases with and without feedback. Specifically, under PBCs, for the former, the system carries  current in the steady state, while the latter  adheres to  detailed balance condition,  resulting in a vanishing current\cite{fengAbsenceLogarithmicAlgebraic2023}. Therefore, it is  reasonable to anticipate distinct relaxation dynamics between the Lindbladians with and without feedback.

Before delving into our findings, we emphasize  the difference of the present work with previous works, which concentrate on the entanglement dynamics by utilizing the quantum jump method to approximately simulate evolution of the system, in terms of the stochastic Schr{\"o}dinger equation\cite{wangAbsenceEntanglementTransition2023,fengAbsenceLogarithmicAlgebraic2023}. In contrast, our work primarily focus on the Liouvillian spectrum, eigenmodes (including steady state) of  Lindbladian and its associated  relaxation dynamics.

\section{Basic knowledge of Lindbladian}
\label{KnowledgeLindbladian}

First of all, let us introduce some basic knowledge of Liouvillian superoperator.
For the fermionic Liouvillian superoperator $\mathcal{L}$,  it is well known
that there is at least one steady state $\rho_{ss}$ if the dimension of the Hilbert space is finite \cite{barthelSolvingQuasifreeQuadratic2022}.
Besides steady state,  there are other (right) eigenmodes $\rho^R_{i}$ satisfying
$\mathcal{L}\rho^{R}_{i}=\lambda_{i}\rho^{R}_{i}$. Physically the real parts of Liouvillian eigenvalues must obey  $\text{Re}[\lambda_{i}]\leq0$. Furthermore, due to  $\mathcal{L}\rho^{R}_{i}=\lambda_{i}\rho^{R}_{i}$, it can readily deduced   $\mathcal{L}{\rho^{R}_{i}}^{\dagger}=\lambda^{*}_{i}{\rho^{R}_{i}}^{\dagger}$, thereby ensuring that the Liouvillian eigenvalues appear in complex conjugate pairs. 
In general, we can order Liouvillian eigenvalues as $0=\text{Re}[\lambda_{0}]>\text{Re}[\lambda_{1}]\geq\text{Re}[\lambda_{2}]\geq ...$, in which  $\lambda_{0}$ represents the unique steady state (in special cases, there exists multiple steady states). The Liouvillian gap is defined as $\Delta=|\text{Re}[\lambda_{1}]|$.
Hence, if the initial density matrix is given by 
$\rho_{\text{ini}}^{R}=\rho_{ss}+\sum_{i}c_i\rho_{i}^{R}$, 
the density matrix at time $t$ can be expressed as $\rho(t)=\rho_{ss}+\sum_{i\neq0}c_{i}e^{\lambda_{i}t}\rho^{R}_{i}$. To preserve the trace, it is evident that  the eigenmodes except for steady state must satisfy $\text{tr}[ \rho_{i}^{R}]=0$. 
Intuitively, the relaxation time will behave as $\tau\sim 1/\Delta$. However, recent works discover  that coefficient $c_i$ can be superexponential, leading to a  discrepancy between the Liouvillian gap and  the relaxation time\cite{moriResolvingDiscrepancyLiouvillian2020,hagaLiouvillianSkinEffect2021}.
 
Alternatively, we can vectorize the Lindblad master equation (see Eq. \ref{LME}) and transform Liouvillian superoperator $\mathcal{L}$ into matrix form $\tilde{\mathcal{L}}$, which is dubbed as Choi-Jamiolkowski isomorphism\cite{CHOI,JAMIOLKOWSKI}. The vectorized master equation is as follows:
\begin{equation}\label{LME2}
	\begin{split}
 \begin{aligned}
\dfrac{d|\rho\rangle\rangle}{dt}&=\tilde{\mathcal{L}}|\rho\rangle\rangle \\
 &=[-i(H\otimes I-I\otimes H^{T}) \\
 &+\gamma\sum_{\mu}(L_{\mu}\otimes L_{\mu}^{*}-\dfrac{1}{2}L_{\mu}^{\dagger}L_{\mu}\otimes I-\dfrac{1}{2}I\otimes L^{T}_{\mu}L^{*}_{\mu})]|\rho\rangle\rangle \\
 &=\left[-i\left(H_{\text{eff}}\otimes I-I\otimes H_{\text{eff}}^* \right)+\gamma\sum_{\mu}L_{\mu}\otimes L_{\mu}^*\right ]|\rho\rangle\rangle,
 \end{aligned}
\end{split}
\end{equation}
 in which $|\rho\rangle\rangle=\sum_{i,j}\rho_{i,j}|i\rangle\otimes|j\rangle$ is the vectorized form of density matrix $\rho=\sum_{i,j}\rho_{i,j}|i\rangle\langle j|$, and the non-Hermitian effective Hamiltonian is given by   $H_{\text{eff}}=H-i\dfrac{\gamma}{2}\sum_{\mu}L_{\mu}^{\dagger}L_{\mu}$. Apparently, $\tilde{\mathcal{L}}$ is a non-Hermitian matrix.
 
 Due to the non-Hermitian nature of Lindbladian $\mathcal{L}$, another set of left  eigenmodes is needed to construct orthogonal relations. The left  eigenmodes are defined as $\mathcal{L}^{\dagger}{\rho_{i}^{L}}=\lambda_{i}^{*}{\rho_{i}^{L}}$,  in which $\mathcal{L}^{\dagger}$ is given by  $\mathcal{L}^{\dagger}\rho=-i[\rho, H]+\sum_{\mu}(\hat{L}_{\mu}^{\dagger}\rho\hat{L}_{\mu}-\dfrac{1}{2}\{\hat{L}_{\mu}^{\dagger}\hat{L}_{\mu}, \rho\})$. It is easy to demonstrate that the right and left
eigenmodes corresponding to different eigenvalues are
orthogonal to each other: $(\rho^{L}_{i}|\rho^{R}_{j})=0\    (\lambda_{i}\neq\lambda_{j})$, where the inner product $(A|B)$ is defined as  $(A|B)=\text{Tr}[A^{\dagger}B]$. Therefore, the coefficients of eigenmodes $\rho_{i}$ can be expressed as $c_{i}=(\rho_{i}^{L}|\rho^{R}_{\text{ini}})/(\rho^{L}_{i}|\rho^{R}_{i})$.

\section{Lindbladian with feedback}
\label{withfeedbcak}

This section is focused on the Lindbladian with feedback, namely ${L}_{i}={U}_{i}{P}_{i}$. It has  been  known that the system carries non-zero current for the feedback case under PBCs  \cite{fengAbsenceLogarithmicAlgebraic2023}, which results in LSE of the steady state under OBCs.  We consider the single-body case in Sec. \ref{feedbacksinglebody} and many-body case in Sec. \ref{feedbackmanybody}.

\subsection{Single-body case}
\label{feedbacksinglebody}
In this subsection, we perform exact diagonalization and perturbation analysis in the single-particle sector and investigate its Liouvillian spectrum $\{\lambda_{i}\}$, steady state $\rho_{ss}$, and Liouvillian gap $\Delta$. Additionally, we analyze its relaxation dynamics, demonstrating the profound impact of LSE on the relaxation behaviors. 

 \begin{figure}[h]
\centering
\includegraphics[height=6.0cm,width=7.5cm]{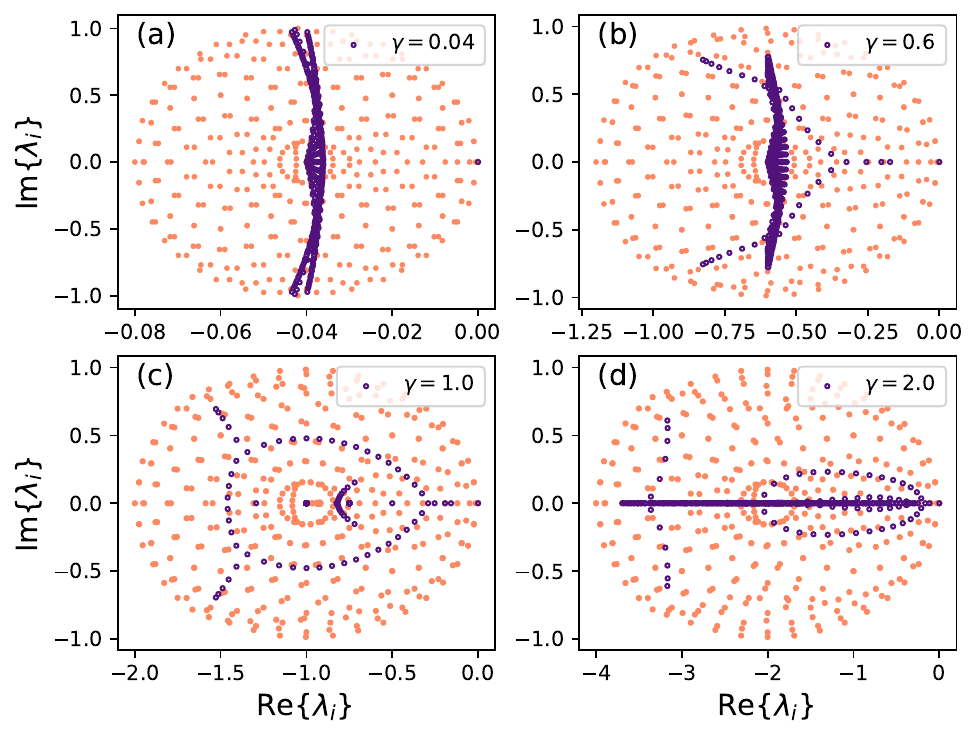}
\caption{Single-particle Liouvillian spectrum for different monitoring rate $\gamma$ under OBCs (purple) and PBCs (orange). The system size is chosen as $L=20$ (a) $\gamma=0.04$, (b) $\gamma=0.6$, (c) $\gamma=1.0$, (d) $\gamma=2.0$. We set precision as 30  digits to avoid numerical inaccuracy.}
\label{fig1}
\end{figure}

\subsubsection{Liouvillian spectrum}
\label{feedbackspectrum}

As show in Fig. \ref{fig1}, the Liouvillian spectrum under PBCs is totally different from OBCs, which is reminiscent of distinct  energy spectrum under PBCs and OBCs for non-Hermitian Hamiltonian hosting NHSE. Indeed, there is deep connection between the Liouvillian spectrum and energy spectrum of the effective non-Hermitian Hamiltonian for the model in this paper, which is directly reflected by the perturbation analysis in Appendix \ref{sec:appendixA}. We will briefly illustrate it in the following. 

For the feedback case under PBCs, we observe that $[H, \sum_{j}L_{j}^{\dagger}L_{j}]=0$, enabling us to take $\tilde{\mathcal{L}}_{0}=-i\left(H_{\text{eff}}\otimes I-I\otimes H_{\text{eff}}^* \right)$ as the unperturbed part. Its corresponding eigenmodes $|k\rangle\otimes|k^{\prime}\rangle^{*}$ are the zeroth-order eigenmodes of  the whole Liouvillian superoperator $\tilde{\mathcal{L}}$, with  $|k\rangle$ denoting Bloch waves with crystal momentum $k$. The resultant zeroth-order eigenvalues are expressed as   $\lambda^{(0)}_{k,k^{\prime}}=-i(E(k)-E(k^{\prime})^{*})=-\dfrac{i}{2}(\text{cos}k-\text{cos}k^{\prime})-\dfrac{\gamma}{2}(2+\text{sin}k+\text{sin}k^{\prime})$. Moreover, the first and second-order corrections of eigenvalues are about $O(L^{-1})$, so for large system size $L$, the border of PBCs  Liouvillian spectrum is well characterized by  $\lambda^{(0)}_{k,k^{\prime}}$. For example, zeroth-order  eigenvalues set the restrictions $-2\gamma\leq\text{Re}\{\lambda_{i}\}\leq 0$ and $-1\leq\text{Im}\{\lambda_{i}\}\leq1$, which agree well with the numerical results. Furthermore, as shown in Fig. \ref{suppfig1} of Appendix. \ref{sec:appendixA}, we also consider the first-order corrections of eigenvalues, which are  more close to the numerical results (see Fig. \ref{fig1}(a)) of exact diagonalization. 

On the other hand, under OBCs, a similar perturbation analysis is performed in Appendix.   \ref{appendixsubsection2}. In contrast with PBCs, the effective non-Hermitian Hamiltonian $H_{\text{eff}}$ exhibits NHSE under OBCs, implying that its eigenstates are exponentially localized instead of extended Bloch waves. Neglecting two terms of unperturbed Liouvillian $\tilde{\mathcal{L}}_{0}$(see Appendix. \ref{appendixsubsection2}),  the zeroth-order Liouvillian eigenvalues roughly  take the form 
\begin{equation}\label{FeedbackOBCSpectrum}
\begin{aligned}
     \lambda^{(0)}_{k,k^{\prime}}&=-i(E_{\text{OBC}}(k)-E_{\text{OBC}}(k^{\prime})^{*}) \\ 
    & \approx\left\{\begin{array}{l}
-\dfrac{i}{2}\sqrt{1-\gamma^{2}}(\text{cos}k-\text{cos}k^{\prime})-\gamma, \text{for}\ \gamma<1  \\
-\dfrac{1}{2}\sqrt{\gamma^2-1}(\text{sin}k+\text{sin}k^{\prime})-\gamma, \text{for}\ \gamma>1,
\end{array}\right.
\end{aligned}
\end{equation}
in which $k\ (k^{\prime})=2\pi j\  (j^{\prime})/L$, $j\  (j^{\prime})=1,2,..L$.
Although the approximation is pretty loosely, $\lambda^{(0)}_{k,k^{\prime}}$ indeed explains some properties of OBCs Liouvillian spectrum. For instance, when $\gamma<1$, as depicted in Fig. \ref{fig1}(a),(b), a large proportion of eigenvalues are located around $\text{Re}[\lambda]=-\gamma$, with the imaginary part of eigenvalues  approximately following  $-\sqrt{1-\gamma^2}\leq\text{Im}\{\lambda_{i}\}\leq\sqrt{1-\gamma^2}$. Moreover, when $\gamma=1$, the eigenvalue $\lambda=-\gamma$ is  hugely degenerate. As for $\gamma>1$, most eigenvalues tend to reside on the real axis.
These observations can be directly interpreted to some extent using  Eq. \ref{FeedbackOBCSpectrum}. However, there are many features of OBCs Liouvillian spectrum  beyond the scope of zeroth-order approximation (i.e. Eq. \ref{FeedbackOBCSpectrum}).
Therefore, as shown in Fig. \ref{suppfig1}(c), we also include the first-order corrections of eigenvalues, which already roughly provide correct shape of Liouvillian spectrum displayed in Fig. \ref{fig1}. Furthermore, it is reasonable to predict that higher corrections will further improve the accuracy.  
In summary, our perturbation analysis reveals that the sensitivity of the Liouvillian spectrum is originated from the sensitivity of the energy spectrum of  the non-Hermitian effective Hamiltonian.

\subsubsection{steady state}
\label{liouvillianskineffect}

As proved in Appendix  \ref{appendixsubsection1}, under PBCs, the steady state can be exactly determined as 
\begin{equation}\label{FeedbackPBCSteady}
\begin{aligned}
\rho_{ss}=\sum_{k}\lambda_{k}|k\rangle\otimes|k\rangle^{*},
\end{aligned}
\end{equation}
 in which $\lambda_{k}=\dfrac{1-\text{sin}k}{1+\text{sin}k}/\sum_{k}\dfrac{1-\text{sin}k}{1+\text{sin}k}$ and $k=\dfrac{2\pi j}{L}, j=1,2,...L$. This steady state is nontrivial, since it is highly coherent. In particular, for $L=4N, N\in\mathcal{Z}$, $\rho_{ss}=\dfrac{1}{L}\sum_{n,m=1}^{L}e^{i\dfrac{\pi}{2}(n-m)}|n\rangle\otimes|m\rangle$, which means the absolute value of all elements of $\rho_{ss}$ is equal to $1/L$. However, it is common for an open quantum system to undergo  decoherence and reach an incoherent steady state finally. Therefore, we attribute this coherent steady state to the intricate interplay between unitary evolution,  continuous measurements and special feedback operations. 

\begin{figure}[h]
\centering
\includegraphics[height=5.5cm,width=7.5cm]{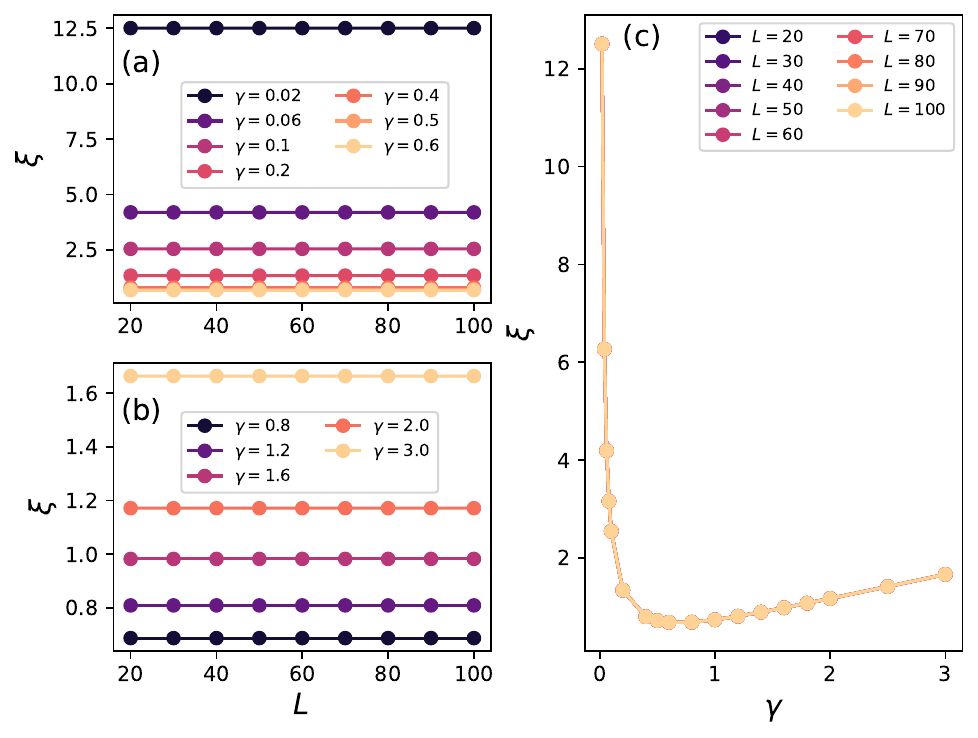}
\caption{Localization length $\xi$ for various $\gamma$ and $L$. (a) $\gamma\leq 0.6$.\ (b)  \ $0.8\leq\gamma\leq 3.0$. (c) The localization length $\xi$ varies with $\gamma$. System sizes are chosen as  $20\leq L\leq 100$. }
\label{fig2}
\end{figure}

As for OBCs, an analytical form of steady state $\rho_{ss}$ is absent. Nevertheless, our  perturbation theory suggests that the majority of Liouvillian eigenmodes will tend to be localized due to the exponential localization behavior of $H_{\text{eff}}$'s eigenstates.
We numerically acquire the steady state under OBCs, whose off diagonal elements are almost zero. Moreover, the diagonal elements exhibits exponential localization behaviors  similar to the Hatano-Nelson model (i.e. $|\rho_{ss}|_{x,x}\propto e^{-x/\xi}$). Therefore,  
we extract the localization length $\xi$ through fitting the diagonal elements of steady state $\rho_{ss}$.
As shown in Fig. \ref{fig2},
the localization length $\xi$ initially experiences a rapid decline with increasing $\gamma$ when $\gamma <0.6$. However, once $\gamma >0.6$, the localization length $\xi$ smoothly increases with growing $\gamma$.
Furthermore, as depicted in Fig. \ref{fig2},  the localization length is independent of system size $L$.
These observed behaviors closely resemble those exhibited by the Hatano-Nelson model.

\begin{figure}[h]
\centering
\includegraphics[height=5.5cm,width=7.5cm]{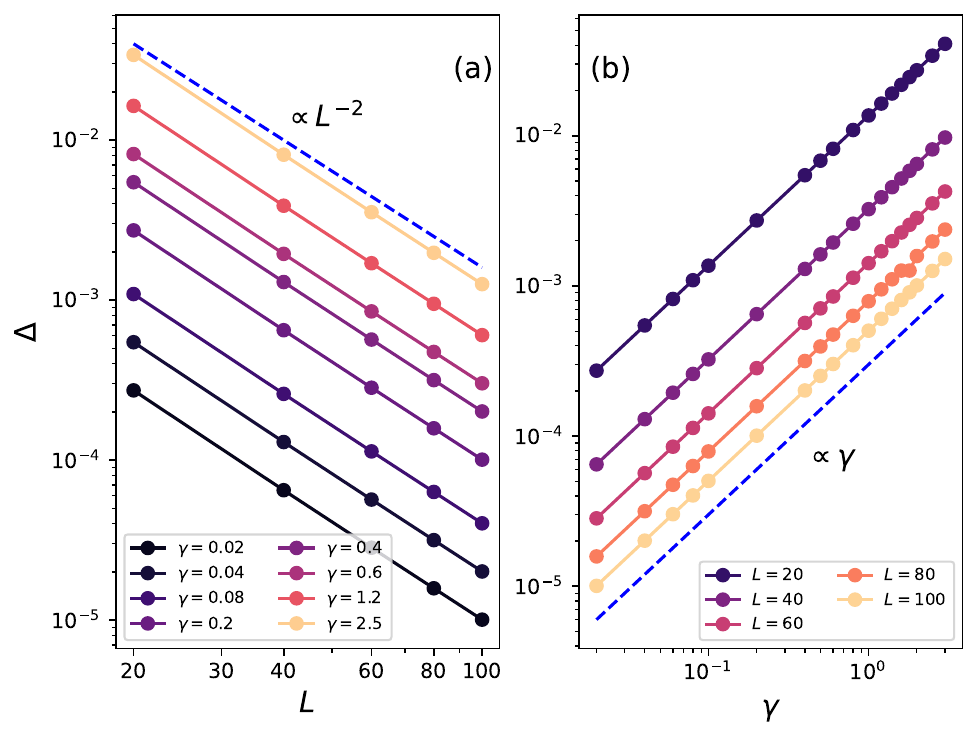}
\caption{Liouvillian gap $\Delta$ versus system size $L$ and monitoring rate $\gamma$ on a log-log plot under PBCs. (a) Liouvillian gap $\Delta$ approximately behaves as $\Delta \propto L^{-2}$. (b) Liouvillian gap  $\Delta$ approximately behaves as $\Delta \propto\gamma$.}
\label{fig3}
\end{figure}

\begin{figure}[h]
\centering
\includegraphics[height=5.5cm,width=7.5cm]{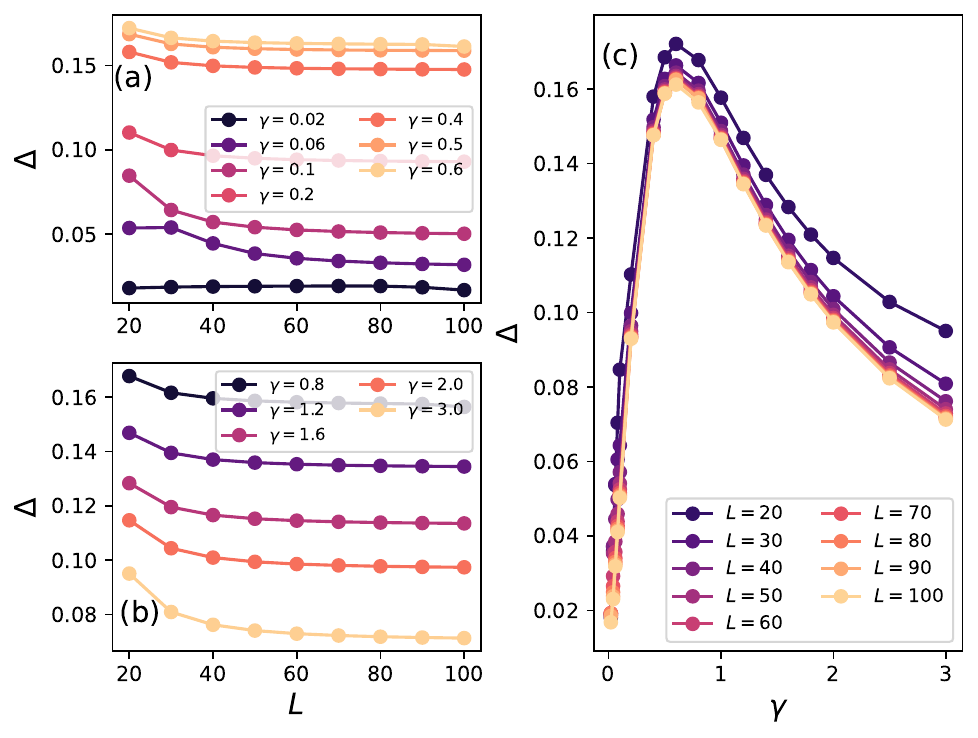}
\caption{Liouvillian gap $\Delta$ for various system size $L$ and monitoring rate $\gamma$ under OBCs. (a) $\gamma\leq0.6$. (b) $0.8\leq\gamma\leq3.0$. The Liouvillian gap saturates at a finite value with the increase in $L$. (c) The Liouvillian gap  firstly increases with $\gamma$ and then decreases. }
\label{fig4}
\end{figure}

The preceding analysis indicates  that the Liouvillian spectrum and Liouvillian eigenmodes, including the steady state, are greatly altered when changing boundary conditions.
In the subsequent section, we delve into a detailed investigation of the distinct  relaxation processes under both  PBCs  and OBCs.
To quantify relaxation time $\tau$, we evaluate distance  $d(t)=||\rho(t)-\rho_{ss}||_{\text{tr}}$, in which  $||\rho(t)-\rho_{ss}||_{\text{tr}}=\text{Tr}\sqrt{(\rho(t)-\rho_{ss})^{\dagger}(\rho(t)-\rho_{ss})}$. We define relaxation time $\tau$  as the smallest time at which $d(t) < 0.01$.

\subsubsection{Liouvillian gap and relaxation dynamics}
\label{relaxationdynamics}

Before discussing  relaxation time $\tau$,  we firstly calculate the scaling relation between Liouvillian gap $\Delta$ and system size $L$. As illustrated  in Fig. \ref{fig3}, we find good scaling relations  $\Delta\propto L^{-2}$ and $\Delta \propto \gamma$ under PBCs. These scaling relations under PBCs  are  well explained by perturbation theory in Appendix. \ref{appendixsubsection1}, from which the Liouvillian gap approximately satisfies   $\Delta\approx\dfrac{\gamma}{2}\left(1+\text{sin}(-\dfrac{\pi}{2}+\Delta k)\right)\approx\dfrac{\gamma\pi^2}{L^2}$. 
On the contrary, as shown in Fig. \ref{fig4}, for arbitrary $\gamma$, Liouvillian gap $\Delta$ remains finite  with the increase in system size $L$ under OBCs. Additionally, for systems with various size $L$ , with the increase in $\gamma$, Liouvillian gap $\Delta$ initially increases  and then decreases.

\begin{figure}[h]
\centering
\includegraphics[height=5.5cm,width=7.5cm]{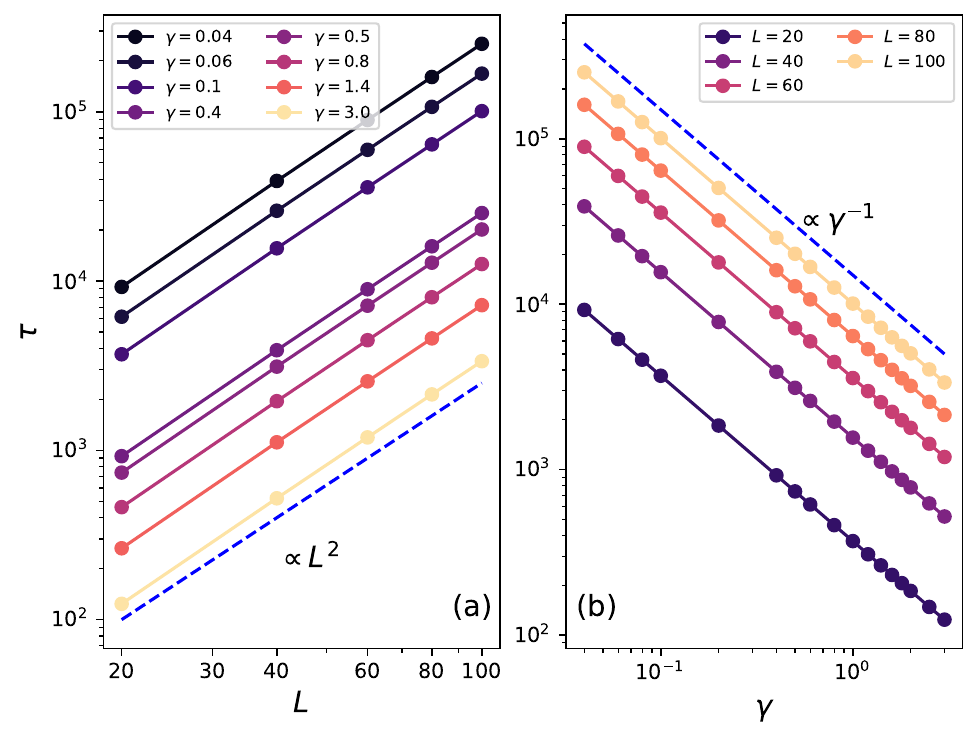}
\caption{Relaxation time versus system size $L$ and monitoring rate $\gamma$ on a log-log plot under PBCs. The initial state is $|L\rangle\langle L|$. The relaxation time $\tau$ is the smallest time $t$ satisfying $d(t)\leq 0.01$. (a) Relaxation time $\tau$ approximately behaves as $\tau\propto L$. (b) Relaxation time $\tau$ approximately behaves as $\tau\propto \gamma^{-1}$.}
\label{fig5}
\end{figure}

\begin{figure}[h]
\centering
\includegraphics[height=5.5cm,width=7.5cm]{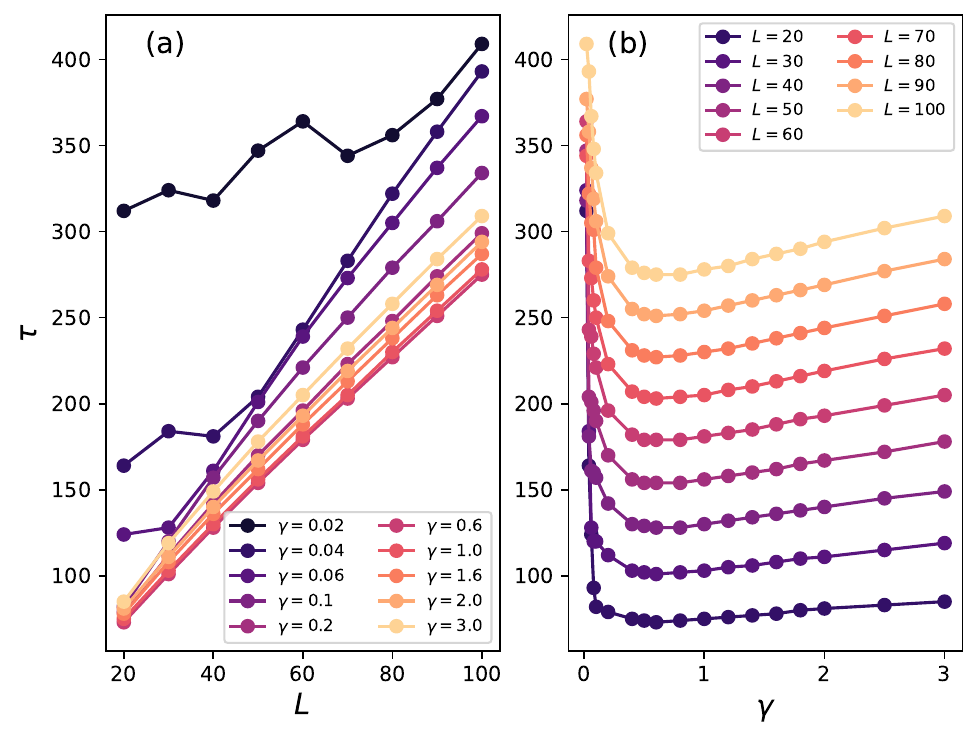}
\caption{Relaxation time $\tau$ for various system size $L$ and monitoring rate $\gamma$ under OBCs. The initial state is $|L\rangle\langle L|$. The relaxation time $\tau$ is the smallest time $t$ satisfying $d(t)\leq 0.01$. (a) For fixed $\gamma$, relaxation time $\tau\propto L.$\  (b)  $\tau$ firstly decreases with $\gamma$ for $\gamma<0.6$ and then slightly increase with $\gamma$.}
\label{fig6}
\end{figure}

As for relaxation time,  we find the relaxation time $\tau$   follows the usual law $\tau \propto 1/\Delta$ under PBCs. Specifically, as indicated  in Fig. \ref{fig5}(a) and (b), the relaxation time $\tau$ exhibits scaling behaviors of  $\tau\propto L^{2}$ and $\tau\propto \gamma^{-1}$,  respectively. Moreover, 
as shown in Fig. \ref{suppfig5}, another evidence is that distance $d(t)$ under PBCs immediately decays as the manner controlled by the inverse of Liouvillian gap, thus inferring $\tau\sim 1/\Delta$.

When switching into OBCs, as shown in Fig. \ref{fig6}(a), the relaxation time is approximately proportional to $L$. However, for small $\gamma$, such as $\gamma=0.02$, $\tau$ deviates from the relation $\tau\propto L$, which is finite-size effect. Because the localization length $\xi$ is comparable with system size $L$ when $\gamma$ is sufficiently small and $L \sim O(10)$. Moreover, as depicted in Fig. \ref{fig6}(b), the  minimum relaxation time occurs around $\gamma=0.6$, at which  the steady state displays  the strongest LSE. Therefore, we can infer that the Liouvillian skin effect can help shorten the relaxation time.
A previous work has unveiled similar scaling relation between relaxation time and system size $L$ ($\tau\sim O(L)$), despite the much difference between our Lindbladians and  theirs\cite{hagaLiouvillianSkinEffect2021}.  This  anomalous  scaling relation under OBCs can be primarily attributed to the  LSE of the first eigenmodes $\rho_{1}^{R}$ and ${\rho_{1}^{R}}^{\dagger}$. The corresponding coefficient follows  $c_{1}=(\rho^{L}_{1}|\rho_{\text{ini}}^{R})/(\rho^{L}_{1}|\rho^{R}_{1})$, in which  $(\rho^{L}_{1}|\rho^{R}_{1})$ is approximately $e^{-O(L/\xi)}$ because of the exponential localization of $\rho^{R}_{1}$ and $\rho^{L}_{1}$ (see Fig. \ref{suppfig2}). Therefore, the relaxation time $\tau$ should approximately obey $e^{L/\xi}e^{-\Delta\tau} \sim e^{-1}$, leading to  $\tau\sim L/\xi\Delta+1/\Delta$\cite{hagaLiouvillianSkinEffect2021}. In the thermodynamic limit, since Liouvillian gap  $\Delta$ is finite, the second term will dominate, yielding $\tau\sim O(L)$. 

\begin{figure}[h]
\centering
\includegraphics[height=5.5cm,width=8.0cm]{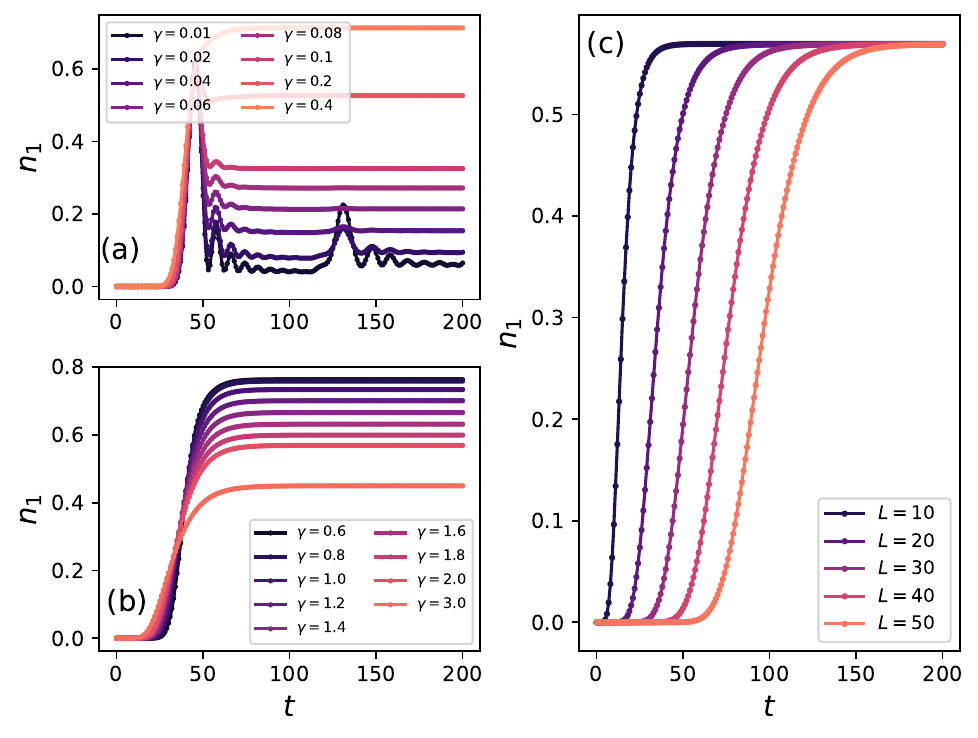}
\caption{Evolution of the first site's density   $n_{1}$ under OBCs. The initial state is $|L\rangle\langle L|$. (a) $L=20$,  $\gamma\leq0.4$. 
(b) $L=20$,  $0.6\leq\gamma\leq3.0$. 
(c) $\gamma=2.0$, $L=10, 20, 30, 40, 50$.}
\label{fig7}
\end{figure}

\begin{figure}[h]
\centering
\includegraphics[height=5.5cm,width=8.0cm]{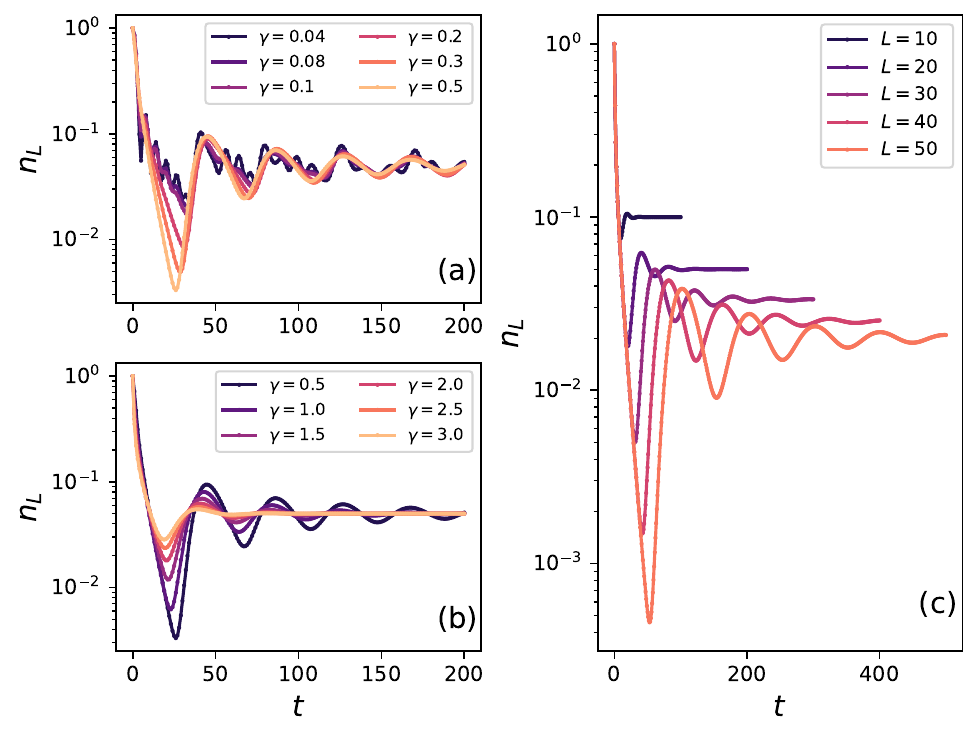}
\caption{Evolution of the last site's density   $n_{L}$ under PBCs on a semi-log plot. The initial state is $|L\rangle\langle L|$. (a) $L=20$, $\gamma\leq 0.5$. (b) $L=20$, $0.5\leq\gamma\leq3.0$. (c) $\gamma=2.0$, $L=10, 20, 30, 40, 50$.}
\label{fig8}
\end{figure}

In addition to distance $d(t)$, we also directly observe the evolution of the density at the first site $\langle n_{1}\rangle$ and the last site $\langle n_{L}\rangle$ to gain  the intuitive pictures for different scaling behaviors of relaxation time. We imagine the initial state $|L\rangle\langle L|$ as a ``particle" situated at site $L$. Under OBCs,  we observe the evolution of density at the first site. When $\gamma > 0.4$, $\langle n_1\rangle$ remains zero until the ``particle" moves to the left,
after which the density rapidly reaches a maximum value and remains constant, which means the ``particle" is sticky to the edge  once the ``particle" touches the boundary. However, for $\gamma < 0.2$, after reaching the peak, $\langle n_1\rangle$  begins  to fall down, suggesting partial reflection of the  ``particle". When $\gamma$ is very small, the evolution of density resembles the unitary case with $\gamma=0$, in which $\langle n_1\rangle$ will constantly irregularly oscillate. For example, for $\gamma=0.01$ or $\gamma=0.02$, it should undergo several oscillations until finally reaches the steady state, so the relaxation time is significantly longer than the big $\gamma$ cases as indicated  in Fig. \ref{fig6}(b). Furthermore, for  $\gamma > 0.4$, it's shown in Fig. \ref{fig7}(b),   the relaxation time for $n_{1}$ is relatively insensitive  to $\gamma$ at least for small sizes, aligning with the results in Fig. \ref{fig6}(b). Moreover, as illustrated in Fig. \ref{fig7}(c), the relaxation time $\tau$ is proportional to system size $L$, which makes sense since the propagation ``velocity" is finite due to the Lieb-Robinson bound\cite{Lieb1972,LiebRobinson2}. Therefore, it always takes at least $O(L)$ time for ``particle" to move through the whole bulk.

In contrast,  the relaxation dynamics under PBCs differs significantly from the case under OBCs.
As depicted in Fig. \ref{fig8}, compared with the OBCs case, the observable $\langle n_{L}\rangle$  undergoes numerous oscillation periods before reaching the steady state.  Notably, larger monitoring $\gamma$  lead to shorter oscillatory periods.  Specifically,  as shown in Fig. \ref{fig8}(c), for $\gamma=2.0$ and $L=10, 20, 30, 40, 50$,  the time of the first peaks of $n_L$ under PBCs are approximately $t\approx20, 42, 62, 83, 103$, which indicates the period  linearly grows $L$. Moreover,  the system should undergo about $N$ periods before reaching the steady state.  We observe $N$ is approximately proportional to $L$, so the relaxation time under PBCs follows $\tau\sim O(L^2)$.  

\begin{figure}[h]
\centering
\includegraphics[height=5.5cm,width=8.0cm]{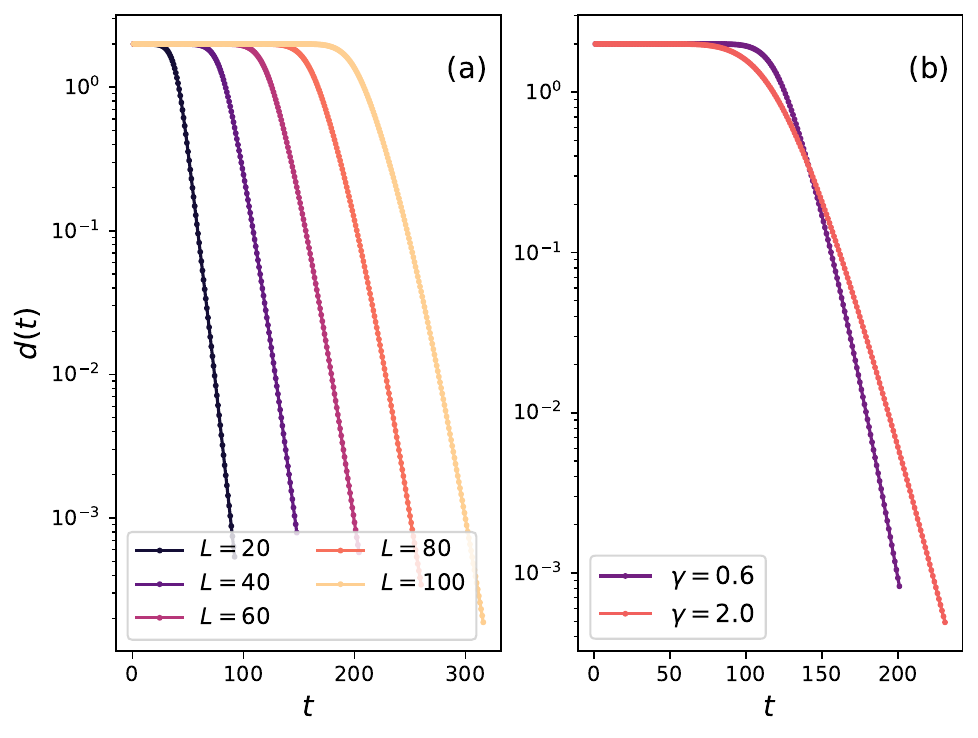}
\caption{Evolution of distance $d(t)$ under OBCs on a semi-log plot.  The initial state is $|L\rangle\langle L|$. (a) $\gamma=0.6$. Distance keeps invariant for $O(L)$ time and then exponentially decays. (b) $L=60$. The LSE for $\gamma=0.6$ is stronger than $\gamma=2.0$, so the cut off phenomenon for the former is more evident. }
\label{fig9}
\end{figure}

Interestingly, for the single-particle case under OBCs, we observe the cutoff phenomena, indicating that relaxation does not occur until a certain time, after which it rapidly proceeds with in a  time window of $O(1)$.   To be concrete, as depicted in Fig. \ref{fig9}, distances $d(t)$ remain constant up to a considerable time ($\sim O(L)$) before  rapidly decreasing.
This phenomenon can be attributed to the steady state $\rho_{ss}$ being close to $|1\rangle\langle1|$, while the initial state is $|L\rangle\langle L|$, ensuring the state $\rho(t)$ ``far from" the steady state $\rho_{ss}$ for a considerable time during evolution.
A straightforward calculation is that  $d(t)\approx\text{tr}\sqrt{(\rho(t)-|1\rangle\langle1|)(\rho(t)-|1\rangle\langle1|)}\approx\text{tr}\sqrt{\rho(t)^2+|1\rangle\langle1|}\approx\text{tr}[\rho(t)+|1\rangle\langle1|]=2$. The approximation relies on the fact that   if the initial state is away from $|1\rangle\langle1|$, then $\rho(t)|1\rangle\langle1|$ and $|1\rangle\langle1|\rho(t)$ will keep close to zero for a while, as $\rho(t)$ takes time to gradually approach $|1\rangle\langle1|$.
Hence, it is clear that cutoff phenomena  depend on initial state. For instance, given a uniform initial state, then  $\rho(t)|1\rangle\langle1|\neq0$, implying that cutoff phenomenon will disappear. Additionally, as depicted in Fig. \ref{fig9}(b), the duration for which the distance remains constant increases with the strength of the skin effect.
Conversely, under PBCs, the cutoff phenomenon is absent due to the delocalized structure of $\rho_{ss}$.   Moreover, cutoff phenomenon provide us another viewpoint to intuitively understand the scaling behavior $\tau\sim 1/\Delta+ L/\xi\Delta$,  which can be divided into two parts $\tau=\tau_1+\tau_2$. As depicted in Fig. \ref{fig9}, the first part is that distance keeps unchanged for a duration $\tau_{1}\sim O(L/\xi\Delta)$, and the second part is that distance roughly exponentially decays as the inverse of Liouvillian gap, thus resulting in $\tau_{2}\sim O(1/\Delta)$.

\subsection{Many-body case  }
\label{feedbackmanybody}
In this section, we  briefly explore the Lindbladian with feedback under OBCs in the many-body half-filled sector. The aim is to see the relaxation behaviors of many-body LSE.

Due to exponential growth of dimensions of the Liouvillian superoperator for the many-body case, the exact diagonalization is limited to  very small system size $L$. Therefore, besides exact diagonalization, we also employ the quantum jump method to simulate the evolution, enabling us to obtain the evolution of the density distribution. The numerical implementations have been elucidated in previous works\cite{quantumtrajectory,wangAbsenceEntanglementTransition2023,fengAbsenceLogarithmicAlgebraic2023}.

\subsubsection{Many-body Liouvillian skin effect}

\begin{figure}[h]
\centering
\includegraphics[height=6.0cm,width=8.0cm]{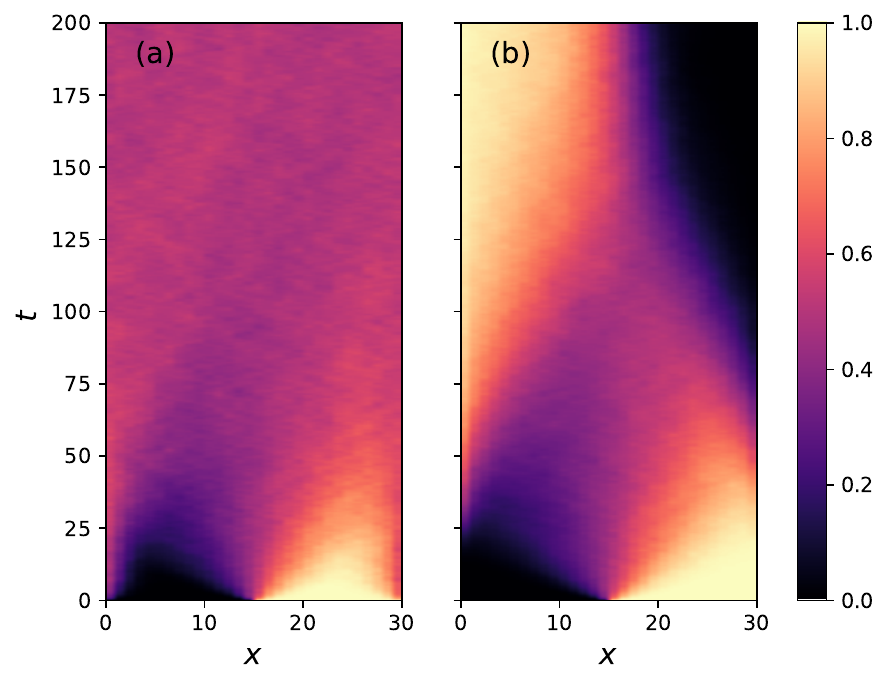}
\caption{Evolution of density distributions. The initial state is  half-filled $|00...011...1\rangle\langle00...011...1|$. $L=30$, $\gamma=2.0$. (a)  PBCs, the density distribution of steady state is uniform. (b) OBCs, the steady state displays skin effect.}
\label{fig10}
\end{figure}

\begin{figure}[h]
\centering
\includegraphics[height=6.0cm,width=8.0cm]{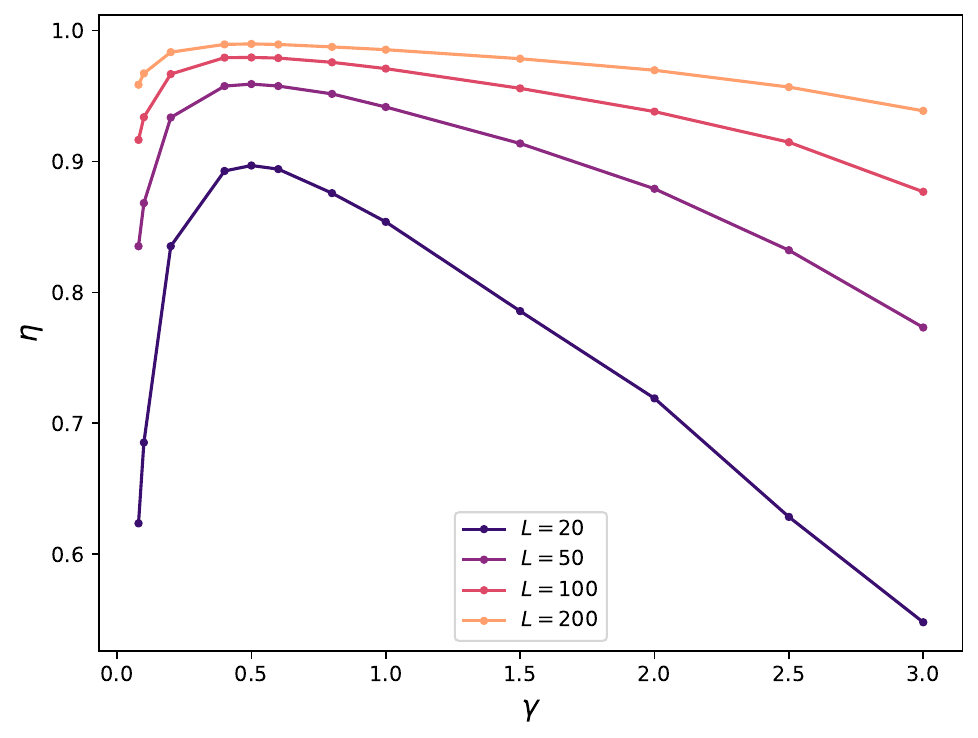}
\caption{Imbalanced density distribution of steady state for many-body half-filled cases with feedback. }
\label{fig11}
\end{figure}

As shown in Fig. \ref{fig10}(b), the steady state under OBCs exhibits the many-body LSE,  which represents that almost all of particles are localized at the left half side, leaving the right half side nearly empty. Previous studies have proved that many-body LSE  immensely suppress the entanglement growth, driving the system into area law phase for any nonzero $\gamma$. This effect arises because most of the particles become frozen due to the Pauli exclusion principle, while only fluctuation areas around the middle contribute to the generation of entanglement\cite{wangAbsenceEntanglementTransition2023,fengAbsenceLogarithmicAlgebraic2023}.

For many-body cases, to quantify the strength of skin effect, we define
\begin{equation}\label{imbalance}
    \eta=\dfrac{N_{\text{left}}-N_{\text{right}}}{N_{\text{tot}}}, 
\end{equation}
 where $N_{\text{left}}$ and $N_{\text{right}}$ represent the number of particles on the left and right halves, respectively.  Obviously, the larger the $\eta$ is, the stronger the skin effect is.  The density distribution is uniform for $\eta=0$, while  particles are entirely localized on the left side for $\eta=1$.

Interestingly, the many-body LSE bears some resemblance to the single-body LSE. Specifically, as depicted in the Fig. \ref{fig11}, $\eta$ is maximum when $\gamma$ is around 0.5, in which the localization length $\xi$ is approximately minimum for the single-body case. In other words, $\gamma\approx0.5\ (\pm0.1)$ is roughly corresponding to the strongest LSE  for both single-body and many-body cases.  In addition, when $\gamma$ is small, the slope is steeper, whereas for larger $\gamma$, the slope is smoother, agreeing with the behavior observed in the single-body case.
However, for the many-body case, $\eta$ will increase with system size $L$ and we suspect that $\eta$ will always equal one in the thermodynamics's limit. We argue that it is because the middle fluctuation zone is $O(1)$, thus the proportion of frozen areas, namely completely occupied or unoccupied zone,  increases with $L$\cite{fengAbsenceLogarithmicAlgebraic2023}. The corresponding single-body picture is that localization length $\xi$ is independent of $L$, so the particle will always be  completely localized in the left half side for sufficiently large $L$. 

\subsubsection{Relaxation dynamics}

\begin{figure}[h]
\centering
\includegraphics[height=4.2cm,width=8.4cm]{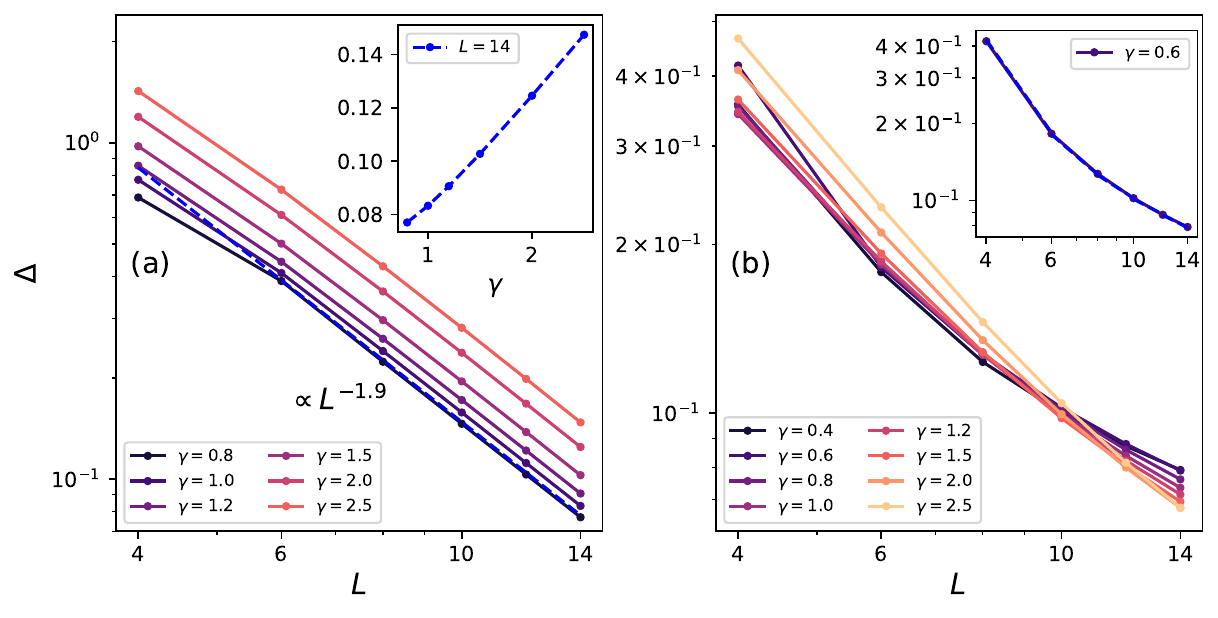}
\caption{Liouvillian gap versus $L$ on a log-log scale for the half-filled case with feedback. (a) PBCs,  Liouvillian gap approximately scales as $\Delta \propto L^{-1.9}$. The inset lists data of  Liouvillian gap for various $\gamma$ when $L=14$. It indicates $\Delta$ linearly grows with $\gamma$. (b) OBCs. For fixed $\gamma$, Liouvillian gap $\Delta$ decreases more and more slowly with the increase in $\gamma$. The inset lists data of Liouvillian gap for  $\gamma=0.6$, the blue dotted line is fitting curve, whose segments are respectively corresponding to $L^{-2.05}, L^{-1.25}$, $L^{-0.93}$, $L^{-0.75}$, $L^{-0.72}$.  }
\label{fig12}
\end{figure}

\begin{figure}[h]
\centering
\includegraphics[height=6.0cm,width=8.0cm]{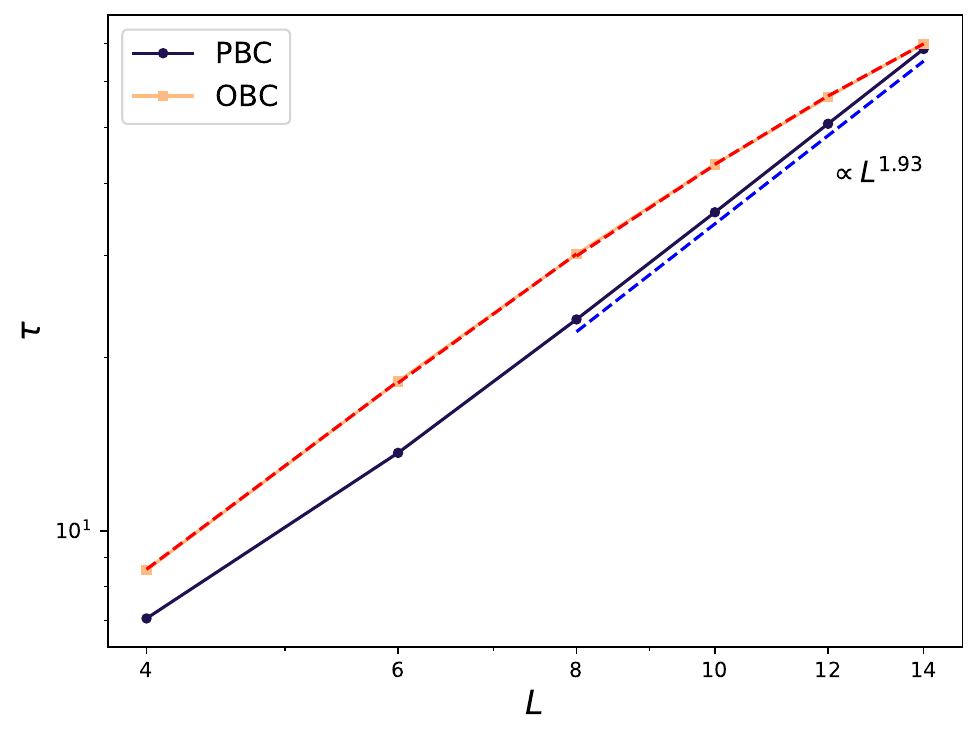}
\caption{Relaxation time $\tau$ versus $L$ on a log-log scale for many-body half-filled case with feedback. The monitoring rate is $\gamma=0.8$. The dashed red line is fitting curve of OBCs data.  The slopes are respectively $\tau\propto$ $L^{1.85}$, $L^{1.79}$, $L^{1.63}$, $L^{1.5}$, $L^{1.35}$. The relaxation time under OBCs is about $\tau\propto L^{1.93}$. It is  worth noting that the relaxation time $\tau$ under PBCs is defined as the smallest time satisfying $d(t)<0.015$, while for OBCs, the $\tau$ is defined as the smallest time obeying $d(t)<0.43$.}
\label{fig13}
\end{figure}

Firstly, we  perform exact diagonalization for the half-filled case up to $L=14$. As Fig. \ref{fig12}(a) shows, under PBCs,  we find the Liouvillian gap approximately scale as  $\Delta\propto L^{-1.9}$ and displays a nearly linear growth with $\gamma$. In contrast,  under OBCs, the Liouvillian gap $\Delta$ decreases with the system size $L$, along with the  decreasing slope. Consequently, it is reasonable to predict that Liouvillian gap $\Delta$ under OBCs will keep finite in the thermodynamic limit. In conclusion, the many-body finite-size results of Liouvillian gap are consistent with the single-body results.

\begin{figure}[h]
\centering
\includegraphics[height=6.0cm,width=8.0cm]{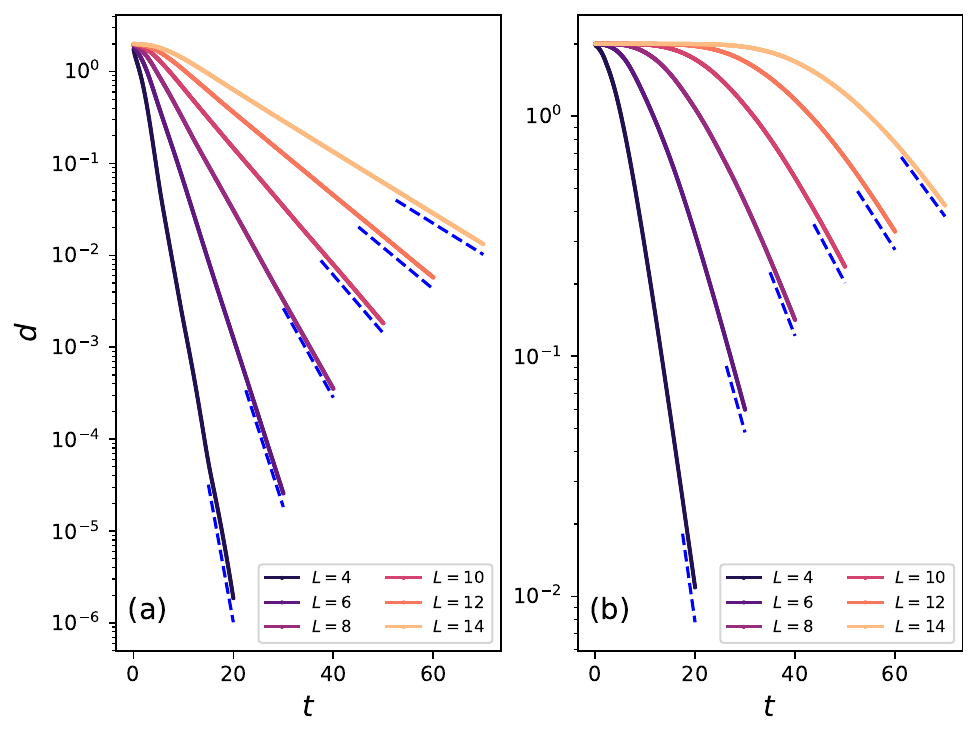}
\caption{Evolution of distance $d(t)$ on a semi-log scale plot for half-filled many-body case with feedback. The monitoring rate is $\gamma=0.8$. (a) PBCs. The distance immediately decays as the inverse of Liouvillian gap. (b) OBCs. Cutoff phenomenon survives in many-body case. }
\label{fig14}
\end{figure}

As shown in Fig. \ref{fig13}, the relaxation time under PBCs  roughly obeys $\tau\propto L^{1.93}$, which approximately satisfies $\tau\sim 1/\Delta$. However, this  simple relation between relaxation time $\tau$ and Liouvillian gap $\Delta$ fails under OBCs, analogous to the single-particle case. As Fig. \ref{fig13} shows, the exponent $\epsilon$ in  $\tau\propto L^{\epsilon}$ under OBCs  constantly  decreases with $L$ within the range of $L$ that we are capable to calculate. We suspect that the relaxation time will still behave as $\tau\sim O(L)$ in the thermodynamic limit. Furthermore,  Fig \ref{fig14} indicates that the cutoff phenomenon still persists in the many-body context under OBCs. Although the Liouvillian gap gives the correct asymptotic decay rate (blue dotted line), the crossover time to get into the   asymptotic regime diverges with system size $L$,  as shown in Fig. \ref{fig14}(b).

\section{Lindbladian Without Feedback}
\label{nofeedback}
In the above section, we have already studied the Lindbladian with feedback, which supports a current-carrying steady state under PBCs and exhibits LSE in the steady state under OBCs. In this section, we  turn to investigate the  Lindbladian without feedback, namely Lindblad operators are given by $\hat{L}_{i}=\hat{P}_{i}$. Without feedback, the Lindblad operators are Hermitian, so the  maximally mixed state $\Bbb{I}/L$ must be one steady state. However, the situation seems more intricate under PBCs. As proved in Appendix. \ref{appendixsubsection3}, when the system size $L=4N, N\in\mathcal{Z}$, there exists bistable steady states, one is $\Bbb{I}/L$, the other is $|-\dfrac{\pi}{2}\rangle\otimes|-\dfrac{\pi}{2}\rangle^*$. It is evident that any linear combinations of $\Bbb{I}/L$ and $|-\dfrac{\pi}{2}\rangle\otimes|-\dfrac{\pi}{2}\rangle^*$ is  steady state as well. In general, bistable steady states can significantly  influence the relaxation process. For example, given different initial states, the system can evolve into different steady states. For simplicity, we will not discuss the dynamical effect caused by  bistable steady states.
In comparison, under OBCs, the steady state is just the maximally mixed state $\Bbb{I}/L$ regardless of system size $L$.

\subsection{Sensitive of boundary condition}

\begin{figure}[h]
\centering
\includegraphics[height=6.0cm,width=8.0cm]{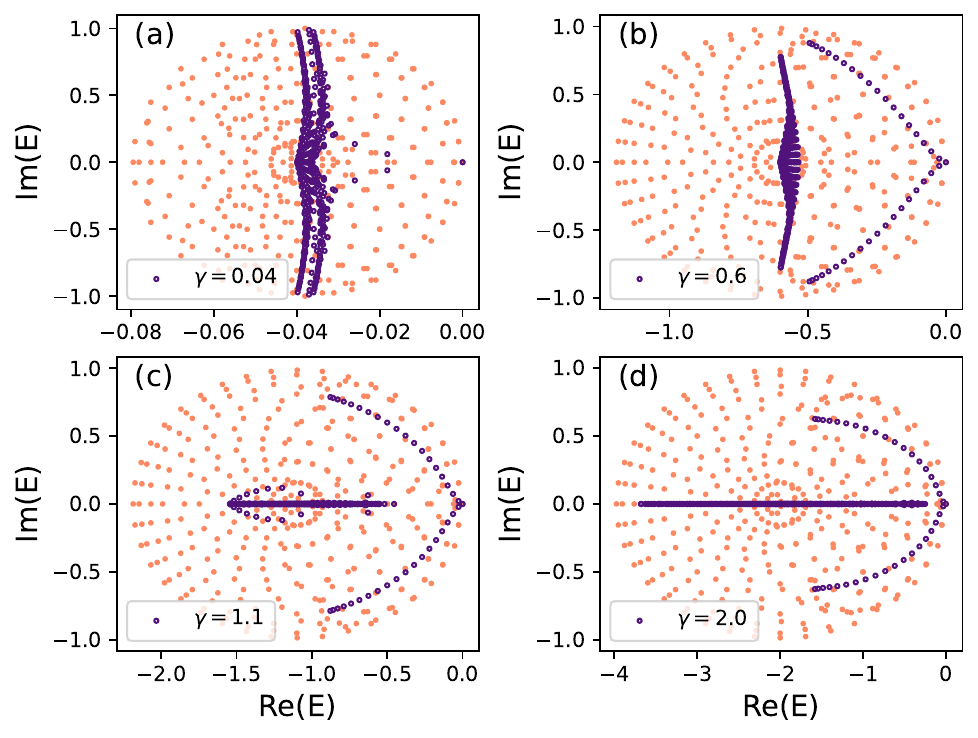}
\caption{Single-particle Liouvillian spectrum versus  different monitoring rate $\gamma$ for no-feedback case under OBCs  (purple) and PBCs (orange). The system size is $L=20$. (a) $\gamma=0.04$, (b) $\gamma=0.6$, (c) $\gamma=1.1$, (d) $\gamma=2.0$. The numerical precision is set to 30 digits to avoid numerical inaccuracy. }
\label{fig15}
\end{figure}

\begin{figure}[h]
\centering
\includegraphics[height=6.0cm,width=8.0cm]{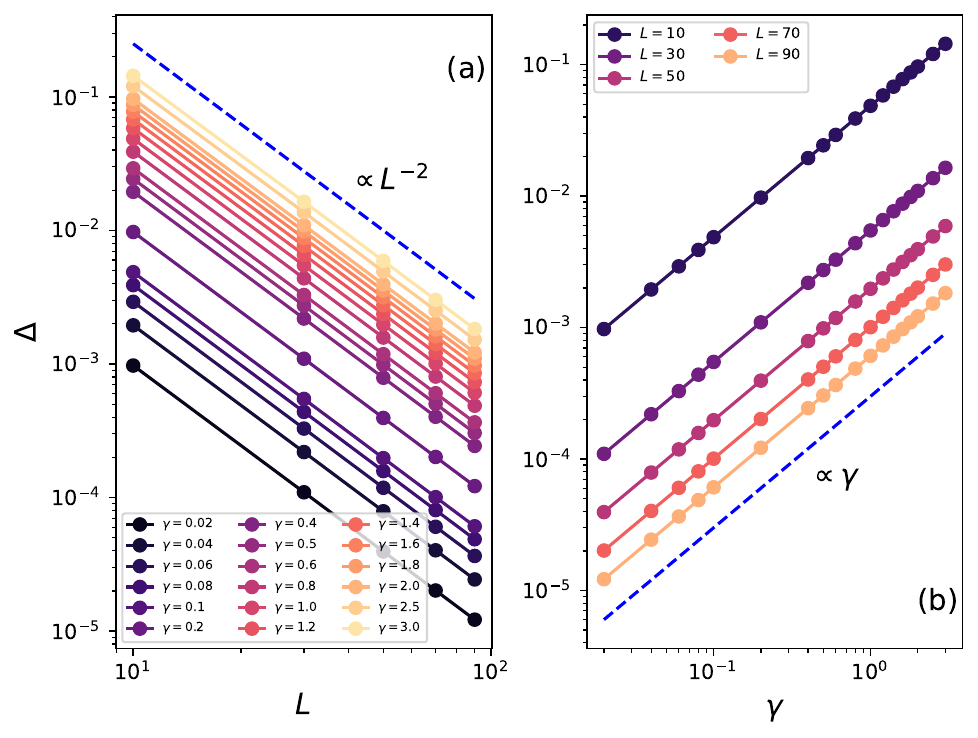}
\caption{Liouvillian gap $\Delta$ versus system size $L$ and monitoring rate $\gamma$ on a log-log plot for no-feedback case under PBCs.  (a) Liouvillian gap  scales  as $\Delta \propto L^{-2}$. (b) Liouvillian gap  $\Delta$ linearly grows with $\gamma$, namely $\Delta \propto\gamma$. }
\label{fig16}
\end{figure}

\begin{figure}[h]
\centering
\includegraphics[height=6.0cm,width=8.0cm]{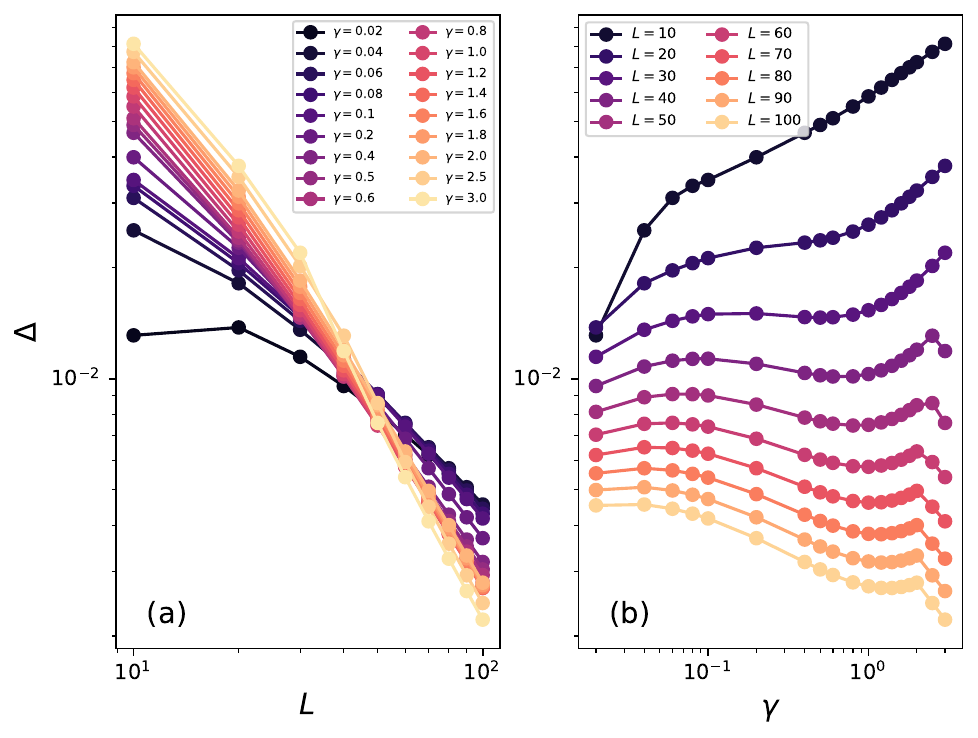}
\caption{Liouvillian gap $\Delta$ versus system size $L$ and monitoring rate $\gamma$ on a log-log plot for no-feedback case under OBCs.  (a) Liouvillian gap $\Delta$ approximately scales as $\Delta \propto L^{-\beta}$, $\beta$ increases with $\gamma$. When $\gamma$ is big,  $\beta\approx2.0$. (b) Liouvillian gap  $\Delta$ seems to be irregular with $\gamma$.}
\label{fig17}
\end{figure}

\begin{figure}[h]
\centering
\includegraphics[height=6.4cm,width=8.6cm]{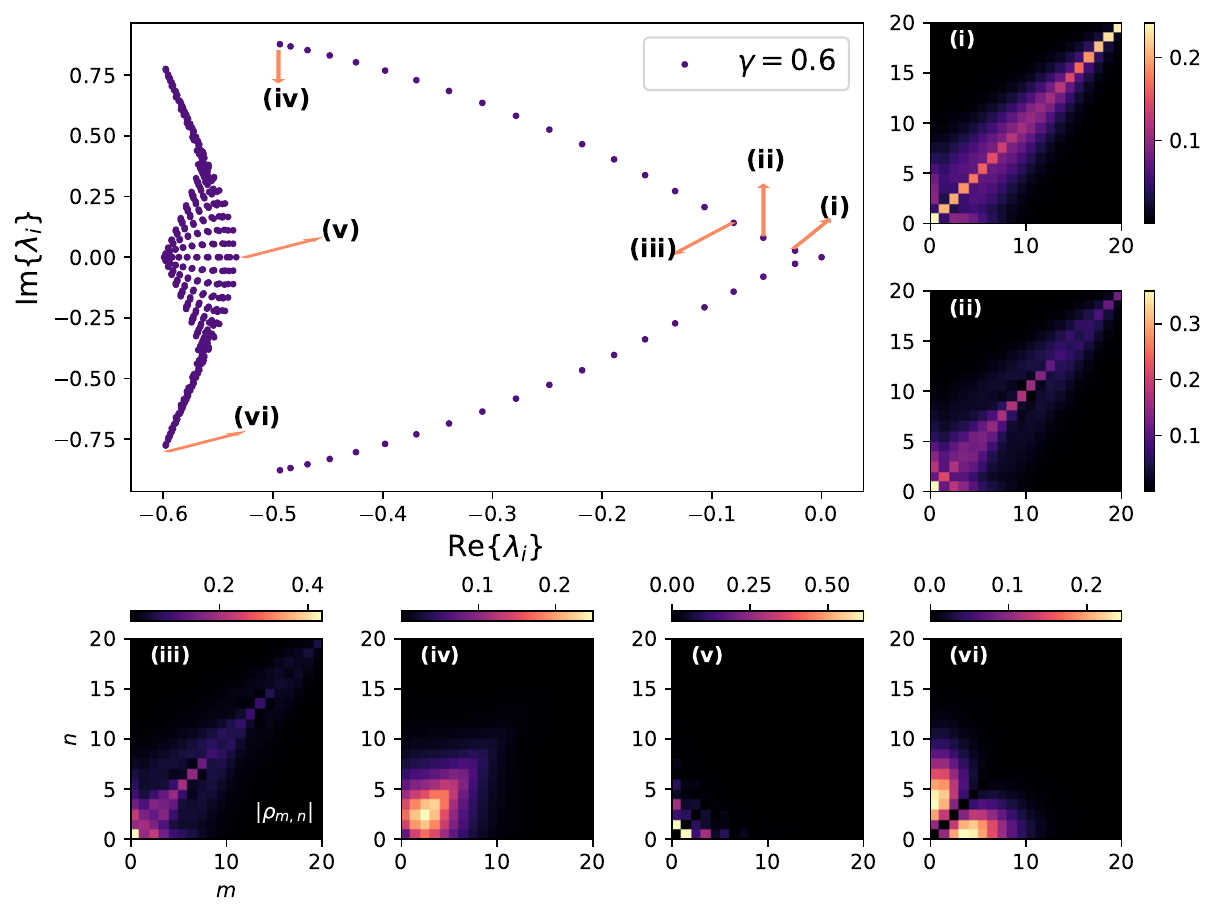}
\caption{Liouvillian spectrum and eigenmodes for no-feedback case under OBCs. The parameters are $L=20, \gamma=0.6$. Apart from  steady state and a few eigenmodes around steady state, the majority of eigenmodes exhibit Liouvillian skin effect. }
\label{fig18}
\end{figure}

Interestingly, even though the steady state is free from skin effect for the no-feedback case, Liouvillian spectrum,  Liouvillian eigenmodes and relaxation dynamics are still sensitive to boundary conditions.

As shown in Fig. \ref{fig15},  the Liouvillian spectrum differs between PBCs and OBCs. The shapes of Liouvillian spectra can also be roughly explained by perturbation theory, similar with the  feedback case (see Appendix. \ref{appendixsubsection3},  \ref{appendixsubsection4} and Fig. \ref{suppfig1}).  Besides the whole Liouvillian spectrum, the Liouvillian gap $\Delta$, namely the real part of first nonzero Liouvillian eigenvalue $\lambda_{1}$, is also sensitive to boundary conditions.  As  shown in Fig. \ref{fig16}, the Liouvillian gap exhibits good scaling behaviors  as $\Delta\propto L^{-2}$ and $\Delta\propto\gamma$ under PBCs, the same with the feedback case under PBCs.
In contrast, as illustrated in Fig. \ref{fig17}, the  Liouvillian gap is irregular under OBCs. Moreover, we find that  $\lambda_{1}$ for no-feedback case under OBCs always has a nonzero imaginary part, which explains the fluctuation of the distance $d(t)$ in Fig. \ref{suppfig6}. 

As for eigenmodes, as indicated in Fig. \ref{fig18}, apart from the  steady state and a few eigenmodes around the steady state, the majority  of Liouvillian eigenmodes exhibit LSE, which is the signal of the sensitivity.  It is consistent with the perturbation analysis, from which we know the zeroth-order solutions display skin effect. Moreover,  we also compare the eigenmodes for no-feedback case under OBCs (see Fig. \ref{fig18}) with the eigenmodes for feedback case under OBCs  (see Fig. \ref{suppfig2}). As a comparison, for the latter, its steady state and eigenmodes around steady states also exhibit strong LSE. This difference leads to completely distinct relaxation behaviors for Lindbladian with feedback and without feedback under OBCs.

\subsection{Relaxation behaviors}

\begin{figure}[h]
\centering
\includegraphics[height=6.0cm,width=8.0cm]{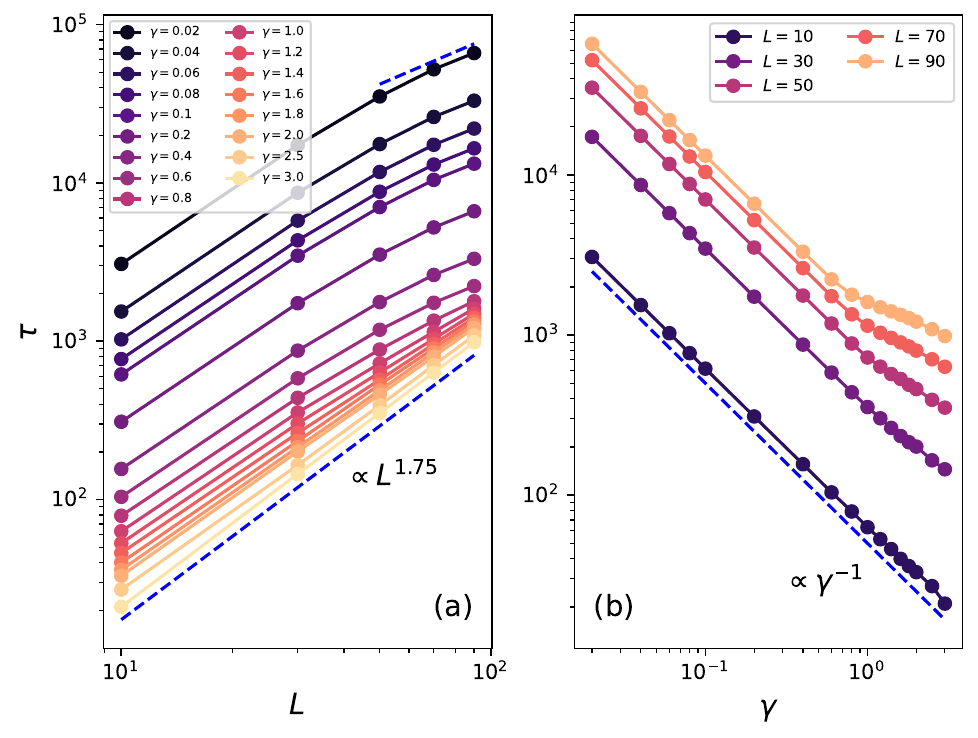}
\caption{Relaxation time $\tau$ versus system size $L$ and monitoring rate $\gamma$ on a log-log plot for no-feedback case under PBCs. The initial state is $|L\rangle\langle L|$. The relaxation time $\tau$ is the smallest time $t$ satisfying $d(t) \leq 0.01$. (a) The relaxation time $\tau\propto L^{\alpha}$, $\alpha<2$. (b)  $\tau$ decreases with the increase in $\gamma$.}
\label{fig19}
\end{figure}

\begin{figure}[h]
\centering
\includegraphics[height=6.0cm,width=8.0cm]{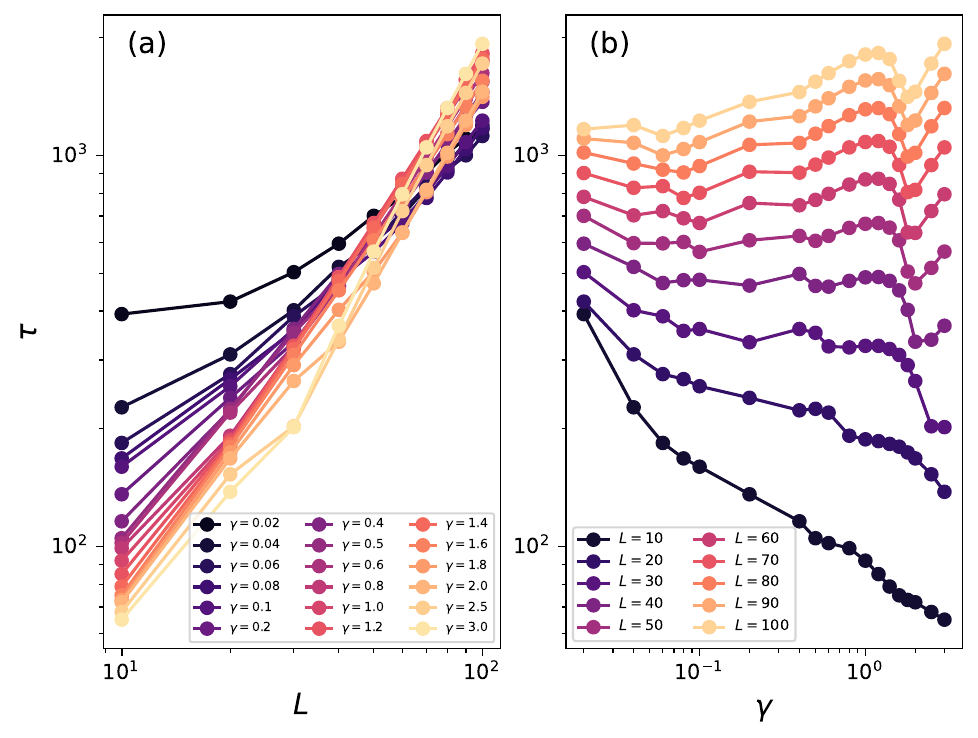}
\caption{Relaxation time $\tau$ versus system size $L$ and monitoring rate $\gamma$ on a log-log plot for no-feedback case under OBCs. The initial state is $|L\rangle\langle L|$. The relaxation time $\tau$ is the smallest time $t$ satisfying $d(t) \leq 0.01$.  (a) The relaxation time $\tau\propto L^{\beta}$, $\beta$ is the same exponent in the scaling relation  $\Delta\propto L^{-\beta}$. (b)  $\tau$ behaves irregularly with $\gamma$. }
\label{fig20}
\end{figure}

\begin{figure}[h]
\centering
\includegraphics[height=6.0cm,width=8.0cm]{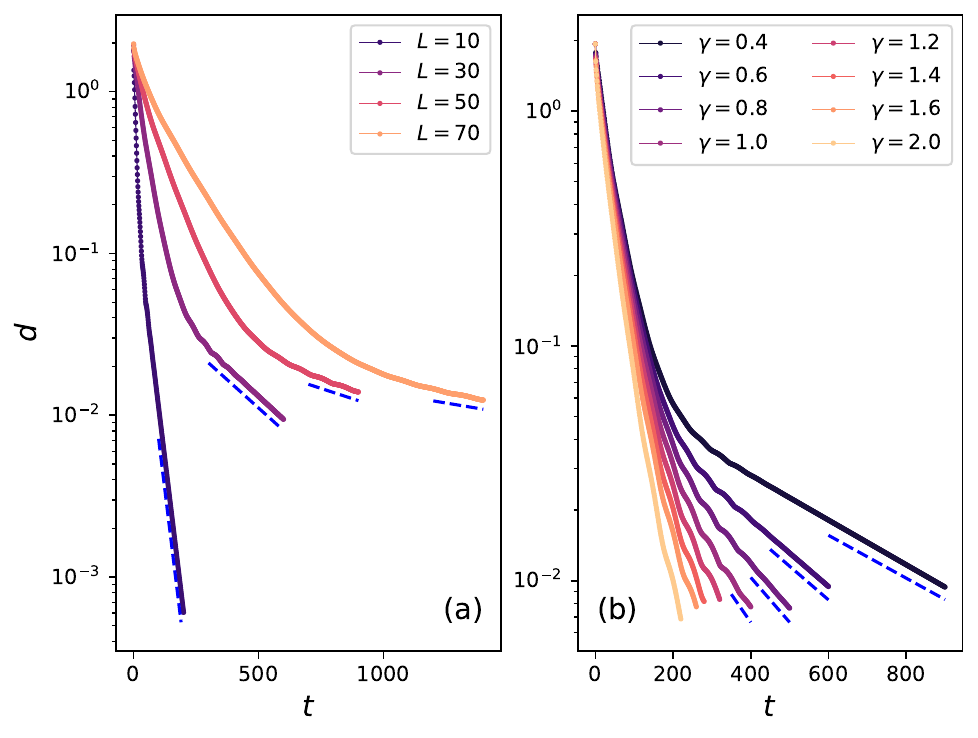}
\caption{Evolution of distance $d(t)$ on a semi-log plot for no-feedback case under PBCs.  Before entering asymptotic regime decay as $e^{-\Delta t}$ (the blue dotted line), the system undergoes faster relaxation. (a) The monitoring rate is  $\gamma=0.6$. (b) System size is $L=30$.}
\label{fig21}
\end{figure}

Due to the sensitivity of Liouvillian spectrum and Liouvillian eigenmodes, it is  natural to expect that the relaxation time is also sensitive to boundary conditions.
Indeed, as shown in Fig. \ref{fig19} and \ref{fig20}, the relaxation time is greatly altered when changing boundary conditions.

Interestingly, for the Lindbladian without feedback under PBCs, we find the relaxation time is much shorter than the inverse of the Liouvillian gap. Specifically speaking, the numerical results (see Fig. \ref{fig19} (a)) indicate $\tau \sim L^{\alpha}, \alpha<2$, while the inverse of the Liouvillian gap gives $\tau\sim O(L^2)$. Moreover, the relaxation time $\tau$ will also deviate from $\tau\propto\gamma^{-1}$ as shown in Fig. \ref{fig19}(b). These phenomenon stems from the accelerated relaxation in the transient regime.
As shown in Fig. \ref{fig21},
before entering the asymptotic regime determined by Liouvillian gap $\Delta$, the system undergoes faster relaxing process for a long time.  This transient
regime takes up a large part for the overall relaxation process, so the faster relaxation rate in the transient regime will greatly decrease the relaxation time and lead to $\tau < 1/\Delta$.
A recent study explains the accelerated relaxation phenomenon in  bulk-dissipative system from the operator spreading, which deserves further research\cite{SymmetrizedLiouvillianGap,shiraiAcceleratedDecayDue2023}. As for the Lindbladian without feedback under PBCs, the faster relaxation has a simple reason. Numerically, we find that  $||c_{1}\rho_{1}^R+c_{1}^{*}{\rho_{1}^{R}}^{\dagger}||\sim O(10^{-2})$, which means other higher eigenmodes, such as $\rho_{2}$, $\rho_{3}$, ..., also play an important role in the relaxation process, thus resulting in   faster decay of distance.  In contrast, for no-feedback case under OBCs, $||c_{1}\rho_{1}^{R}+c_{1}^{*}{\rho_{1}^R}^{\dagger}||\sim O(10^{0})$, implying that the $\rho_{1}$ is the dominant term for relaxations.

For the Lindbladian without feedback under OBCs, we find that  the relaxation time is again controlled by the inverse of Liouvillian gap $\Delta$. Firstly, we have numerically checked  the scaling behavior of relaxation time $\tau$ in Fig. \ref{fig20}, which is well fitted with the inverse of Liouvillian gap in Fig. \ref{fig17} for various $\gamma$.  Another evidence is to observe the evolution of distance $d(t)$ (see Fig. \ref{suppfig6}). We discover that the distance immediately decays in the manner controlled by Liouvillian gap $\text{Re}(\lambda_1)$, and the fluctuation of distance is caused by the imaginary part of $\lambda_{1}$.  Physically, it is reasonable since the steady state and low Liouvillian eigenmodes are free from skin effect. Therefore, the coefficients of eigenmodes around the steady state are still finite,  although the coefficients of high eigenmodes can be exponentially large. After a short time, the relaxation process is dominated by $\rho_{1}^{R}$, and thus leading to $\tau\sim 1/\Delta$. Moreover, the relation between the relaxation time $\tau$ and monitoring rate $\gamma$ under OBCs is non-monotonic, while the relaxation time $\tau$ under PBCs always decreases with  $\gamma$.

\section{Conclusion and Outlook}
\label{sec:conclsn}

In summary, we have investigated the Liouvillian spectrum, eigenmodes, and relaxation dynamics of two Lindbladians, which share the same non-Hermitian effective Hamiltonian.
We find both Lindbladians are sensitive to boundary conditions due to LSE, which is originated from the NHSE of non-Hermitian effective Hamiltonian. When changing the boundary conditions, the Liouvillian spectrum, Liouvillian eigenmodes and relaxation dynamics are all dramatically altered.  

To be specific, we list the main results as follows:  (i) For the Lindbladian with feedback  under PBCs, the steady state is highly coherent and eigenmodes are extended.  The relaxation time is controlled the inverse of the Liouvillian gap $\tau\sim 1/\Delta$ and scales as  $\tau\sim O(L^2)$.
(ii) For the Lindbladian with feedback under OBCs, because the corresponding Lindbladian under PBCs carries current in the steady state, the steady state will be localized under OBCs. In other words, the eigenmodes including the steady state exhibit LSE, which is able to lead to cutoff phenomenon. Moreover, the LSE of low eigenmodes, especially for $\rho_{1}$, can  result in  exponentially large coefficients $c_{1}$. As a consequence, the relaxation time obeys $\tau\sim\dfrac{1}{\Delta}+\dfrac{L}{\xi\Delta}$, implying  $\tau\sim O(L)$ in the thermodynamic limit. Our finite-size results  support that the LSE and distinct  scaling relations between relaxation time $\tau$ and system size $L$ under PBCs and OBCs still hold in many-body system.
(iii)
For the Lindbladian without feedback  under PBCs, we find if system size $L\neq 4N, N\in\mathcal{Z}$, the steady state is the unique maximally mixed state $\Bbb{I}/L$. Interestingly, if the system size $L=4N, N\in\mathcal{Z}$, there exists bistable steady states, one is $\Bbb{I}/L$ and another is $|-\dfrac{\pi}{2}\rangle\otimes|-\dfrac{\pi}{2}\rangle$. Moreover, the eigenmodes are extended due to the translation invariance. Remarkably, we observe that  the relaxation time  ($\tau\sim O(L^{\alpha}), 1<\alpha<2$) is smaller than the inverse of the Liouvillian gap ($\tau\sim L^{2}$), which stems from the faster relaxation persisting for a long time in the transient regime before entering the asymptotic regime governed by the Liouvillian gap. (iv) For the Lindbladian without  feedback under OBCs, although the majority of eigenmodes exhibit LSE,  the steady state and a few eigenmodes around steady state are extended. Moreover, we find the eigenmodes $\rho_{1}$ and $\rho_{1}^{\dagger}$ dominate the decay of the distance immediately, which implies that  the relaxation time follows $\tau\sim 1/\Delta$.  Consequently, the relaxation time
scales as $\tau\propto L^{\beta}$, in which $\beta$ is  the same  scaling exponent of the Liouvillian gap  $\Delta\propto L^{-\beta}$.  Moreover, the relation between relaxation time $\tau$ and monitoring rate $\gamma$ is also dramatically altered when changing boundary conditions. Roughly speaking, for both  Lindbladians with feedback and without feedback under PBCs, the relaxation time $\tau$ decreases with  $\gamma$,  while the  relaxation time $\tau$  presents non-monotonic dependence on $\gamma$ under OBCs.

Our work investigates two typical  $U(1)$ symmetric  boundary-sensitive Lindbladians, whose sensitivity originates from Liouvillian skin effect.
There are several interesting  directions which merit further study. First of all, it remains unclear 
whether the single-particle relaxation behaviors still generally hold in many-body case?
Secondly, it is tempting to construct other boundary sensitive Lindbladians beyond skin effect. For example, the topological edge mode under OBCs can  also dramatically alter the relaxation properties\cite{vernierMixingTimesCutoffs2020}.
Another direction is to investigate the nontrivial dynamics induced by the interplay of the Anderson localization and the skin effect\cite{jiangInterplayNonHermitianSkin2019,liDisorderInducedEntanglementPhase2023}. It is  reasonable to predict the Anderson localization will compete with skin effect and may enrich the relaxation dynamics.
Lastly, our work shows that  Lindbladians with current-carrying steady state  and Lindbladians with no-current steady state exhibit different relaxation behaviors. So, it is  worthwhile to investigate the dynamics of other Lindbladians  hosting current-carrying steady state, such as  asymmetric exclusion process\cite{ASEP}, and uncover possible universal laws.

\begin{acknowledgments}
We thank Shuo, Liu for helpful  discussions.
The work is supported by National Key Research
and Development Program of China (Grant No.
2021YFA1402104), the NSFC under Grants No.12174436
and No.T2121001 and the Strategic Priority Research
Program of Chinese Academy of Sciences under Grant
No. XDB33000000.

\end{acknowledgments}

\bibliographystyle{apsreve}
\bibliography{ref}

\clearpage
\appendix
\renewcommand{\theequation}{S\arabic{equation}}
\setcounter{equation}{0}
\renewcommand{\thefigure}{S\arabic{figure}}
\setcounter{figure}{0}
\section{Perturbation theory of Lindbladians}
\label{sec:appendixA}

\subsection{With feedback, PBCs}
\label{appendixsubsection1}

In this section, we investigate the Lindbladian  with feedback under PBCs. The discussion will be restricted in the  single-particle sector. 
The Hamiltonian is given as $H=\dfrac{1}{4}\sum_{j=1}^{L}(c_{j}^{\dagger}c_{j+1}+c_{j+1}^{\dagger}c_{j})$. Moreover, the Lindblad operators are $L_{j}=\dfrac{1}{2}(n_{j}-n_{j+1})+\dfrac{i}{2}(c^{\dagger}_{j}c_{j+1}+c_{j+1}^{\dagger}c_{j})$ and $L^{\dagger}_{j}L_{j}=\dfrac{1}{2}(n_{j}+n_{j+1})+\dfrac{i}{2}(c^{\dagger}_{j}c_{j+1}-c^{\dagger}_{j+1}c_{j})$. 

\subsubsection{Liouvillian spectrum and gap}
Firstly, it can be easily verified $[H, \sum_{j=1}^{L}L^{\dagger}_{j}L_{j}]=0$, so $H$ and $\sum_{j=1}^{L}L^{\dagger}_{j}L_{j}$ will share common eigenstates. By virtue of this property, we decompose the Lindbladian $\tilde{\mathcal{L}}$ into two parts: $\tilde{\mathcal{L}}=\tilde{\mathcal{L}_{0}}+\gamma\tilde{\mathcal{L}_{1}}$. $\tilde{\mathcal{L}_{0}}$ and $\tilde{\mathcal{L}_{1}}$ are respectively as follows:
\begin{equation}
\begin{aligned}
\tilde{\mathcal{L}_{0}}&=-i\left(H_{\text{eff}}\otimes I-I\otimes H_{\text{eff}}^{*} \right) \\
\tilde{\mathcal{L}_{1}}&=\sum_{j=1}^{L}L_{j}\otimes L_{j}^{*},
\end{aligned}
\end{equation}
in which $H_{\text{eff}}=H-i\dfrac{\gamma}{2}\sum_{j=1}^{L}L^{\dagger}_{j}L_{j}$ and $H_{\text{eff}}^{*}=H^{T}+i\dfrac{\gamma}{2}\sum_{j=1}^{L}L_{j}^{T}L_{j}^{*}$.
For the translation invariant free fermion Hamiltonian $H$, we can perform Fourier transformation $c_{j}=\dfrac{1}{\sqrt{L}}\sum_{k}e^{-ikj}c_{k}$, and thus $H$ can be rewritten as  $H=\sum_{k}\dfrac{1}{2}\text{cos}kc^{\dagger}_{k}c_{k}$, in which $k=\dfrac{2\pi m}{L}$, $m=1,2,...L$. $k_{1}$ and $k_{2}$ are equivalent, if they differ by integer multiples of 2$\pi$. Obviously, $|k\rangle\otimes|k^{\prime}\rangle^{*}$ are right eigenvectors  of the Liouvillian $\tilde{\mathcal{L}}_{0}$, satisfying $\tilde{\mathcal{L}}_{0}|k\rangle\otimes|k^{\prime}\rangle^{*}=\left(-\dfrac{i}{2}(\text{cos}k-\text{cos}k^{\prime})-\dfrac{\gamma}{2}(2+\text{sin}k+\text{sin}k^{\prime}) \right)|k\rangle\otimes|k^{\prime}\rangle^{*}$, namely the zeroth-order eigenvalues are $\lambda_{k,k^{\prime}}^{(0)}=-\dfrac{i}{2}(\text{cos}k-\text{cos}k^{\prime})-\dfrac{\gamma}{2}(2+\text{sin}k+\text{sin}k^{\prime}) $. Utilizing the perturbation theory, the first-order corrections of eigenvalues can be obtained as follows:
\begin{equation}
\begin{aligned}
\lambda^{(1)}_{k,k^{\prime}}&=(\langle k|\otimes \langle k^{\prime}|^{*})\ \tilde{\mathcal{L}_{1}}\ (|k\rangle\otimes|k^{\prime}\rangle^{*})\\
&=\sum_{j=1}^{L}\langle k|L_{j}|k\rangle(\langle k^{\prime}||L_{j}|k^{\prime}\rangle)^{*}\\
&=\sum_{j=1}^{L}(\dfrac{i}{L}\text{cos}k)(\dfrac{-i}{L}\text{cos}k^{\prime}) \\
&=\dfrac{1}{L}\text{cos}k\  \text{cos}k^{\prime}.
\end{aligned}
\end{equation}
Furthermore, the second-order corrections of eigenvalues take the form 

\begin{equation}\label{appendixEq2}
\begin{aligned}
\lambda^{(2)}_{k,k^{\prime}}&=
\sum_{(k_{2},k_{2}^{\prime})\neq(k,k^{\prime})} \\ 
&\dfrac{(\langle k_{2}|\otimes\langle k_{2}^{\prime}|^{*}\ \tilde{\mathcal{L}}_{1}\ |k\rangle\otimes|k^{\prime}\rangle^{*} )(\langle k|\otimes\langle k^{\prime}|^{*} \tilde{\mathcal{L}}_{1}\ |k_{2}\rangle\otimes|k_{2}^{\prime}\rangle^{*})}{\lambda^{(0)}_{k,k^{\prime}}-\lambda^{(0)}_{k_{2},k_{2}^{\prime}}} \\ 
&=\sum_{(k_{2},k_{2}^{\prime})\neq(k,k^{\prime})} 
\dfrac{\left(\sum_{j=1}^{L}L_{j}^{k_{2}k}{L_{j}^{k_{2}^{\prime}k^{\prime}}}^{*}\right)\left(\sum_{l=1}^{L}L_{l}^{kk_{2}}{L_{l}^{k^{\prime} k_{2}^{\prime}}}^{*}\right)}{\lambda^{(0)}_{k,k^{\prime}}-\lambda^{(0)}_{k_{2},k_{2}^{\prime}}}, 
\end{aligned}
\end{equation}
in which ${L_{j}}^{k_{2}k}=\langle k_{2}|L_j|k\rangle$ and ${L_{j}^{k_{2}^{\prime}k^{\prime}}}^{*}=(\langle k_{2}^{\prime}|L_j|k^{\prime}\rangle)^{*}$.
Then substituting $\langle k_{2}|L_{j}|k\rangle=\dfrac{1}{2L}e^{i(k_{2}-k)j}(1-e^{i(k_2-k)}+i e^{-i k}+i e^{i k_2})$ into the above equation, the  Eq. \ref{appendixEq2} is reduced to
\begin{equation}
\begin{aligned}
\lambda^{(2)}_{k,k^{\prime}}&=
\dfrac{1}{L^2}\sum_{k_{2}\neq k}\dfrac{\text{cos}k \text{cos}k_{2} \text{cos}k^{\prime} \text{cos}(k_2+k^{\prime}-k)}{\lambda^{(0)}_{k,k^{\prime}}-\lambda^{(0)}_{k_{2},k_{2}+k^{\prime}-k}}. 
\end{aligned}
\end{equation}
The eigenvalues are approximated as  $\lambda_{k,k^{\prime}}\approx\lambda^{(0)}_{k,k^{\prime}}+\gamma\lambda^{(1)}_{k,k^{\prime}}+\gamma^{2}\lambda^{(2)}_{k,k^{\prime}}$. In general, for relatively large $L$, because both $\lambda^{(1)}_{k,k^{\prime}}$ and $ \lambda^{(2)}_{k,k^{\prime}} $
are about $O(1/L)$, 
$\lambda^{(1)}_{k,k^{\prime}}, \lambda^{(2)}_{k,k^{\prime}} \ll \lambda^{(0)}_{k,k^{\prime}} $. In other words, Liouvillian spectrum is dominated by $\lambda^{(0)}_{k,k^{\prime}}$. Therefore, the real and imaginary part of Liouvillian spectrum are  $\text{Re}[\{\lambda\}]=\{-\dfrac{\gamma}{2}(2+\text{sin}k+\text{sin}k^{\prime})|k,k^{\prime}\in[-\pi,\pi)\}$ and $\text{Im}[\{\lambda\}]=\{-\dfrac{1}{2}(\text{cos}k-\text{cos}k^{\prime})|k,k^{\prime}\in[-\pi,\pi) \}$, respectively, which is consistent with  Fig. \ref{fig1}, which shows $-2\gamma\leq\text{Re}[\{\lambda\}]\leq0$ and $-1\leq\text{Im}[\{\lambda\}]\leq 1$.

The first-order perturbation of the eigenvectors is 
\begin{widetext}
\begin{equation}
\begin{aligned}
&(|k\rangle\otimes|k^{\prime}\rangle^{*})^{(1)}=\sum_{(k_2,k^{\prime}_{2})\neq(k,k^{\prime})}\dfrac{\langle k_{2}|\otimes\langle k^{\prime}_{2}|^{*}\ \tilde{\mathcal{L}_{1}}|k\rangle\otimes|k^{\prime}\rangle^{*}}{\lambda^{(0)}_{k,k^{\prime}}-\lambda^{(0)}_{k_2,k^{\prime}_{2}}}|k_{2}\rangle\otimes|k^{\prime}_{2}\rangle^{*}\\
&=\dfrac{1}{2L}\sum_{k_{2}\neq k} \dfrac{1+\text{sin}k+\text{sin}k^{\prime}-\text{sin}k_{2}+\text{cos}(k-k^{\prime})+\text{cos}(k_2+k^{\prime})-\text{cos}(k_2-k)-\text{sin}(k_2+k^{\prime}-k)}{\lambda^{(0)}_{k,k^{\prime}}-\lambda^{(0)}_{k_2,k_2+k^{\prime}-k}}|k_{2}\rangle\otimes|k^{\prime}_{2}\rangle^{*}.
\end{aligned}    
\end{equation}
\end{widetext}
Obviously, the first-order correction of eigenvector  $(|-\dfrac{\pi}{2}\rangle\otimes|-\dfrac{\pi}{2}\rangle^{*})^{(1)}=0$. Moreover, the zeroth-order and higher order eigenvalues for $|-\dfrac{\pi}{2}\rangle\otimes|-\dfrac{\pi}{2}\rangle^{*}$ satisfy  $\lambda_{-\pi/2,-\pi/2}^{(0)}=\lambda_{-\pi/2,-\pi/2}^{(1)}=\lambda_{-\pi/2,-\pi/2}^{(2)}=0$. Therefore, it's straightforward to speculate that   $|-\dfrac{\pi}{2}\rangle\otimes|-\dfrac{\pi}{2}\rangle^{*}$ is the steady state for the whole Lindbladian  $\tilde{\mathcal{L}}$ and indeed we can readily verify it later. 

The first and second-order  perturbation analysis above is restricted in non-degenerate  case. Besides $L$ non-degenerate zeroth-order eigenvalues, there are also $L^{2}-2L$ and  $L$ zeroth-order eigenvalues, whose degeneracies are  2 and  $L$, respectively. For instance, $\lambda_{k,2\pi-k}^{(0)}$ is $L$-fold degenerate, and $\lambda_{\pi,\pi-2\pi/L}^{(0)}$ is two-fold degenerate (degenerate with $\lambda_{2\pi/L,0}$).
According to degenerate perturbation theory, assume $|k_{1}\rangle\otimes|k^{\prime}_{1}\rangle^{*}$ is degenerate with $|k_{2}\rangle\otimes|k^{\prime}_{2}\rangle^{*}$, the corresponding first-order correction of eigenvalues is given by
\begin{equation}\label{}
\begin{aligned}
    \lambda^{(1)}_{\alpha} &= \dfrac{1}{2}\left[\tilde{\mathcal{L}}_{1}^{\alpha_{1}\alpha_{1}}+\tilde{\mathcal{L}}^{\alpha_{2}\alpha_{2}} \right. \\ 
    & \left. \pm\sqrt{(\tilde{\mathcal{L}_{1}}^{\alpha_{1}\alpha_{1}}-\tilde{\mathcal{L}_{1}}^{\alpha_{2}\alpha_{2}})^2+4\tilde{\mathcal{L}}_{1}^{\alpha_{1}\alpha_{2}}\tilde{\mathcal{L}}_{1}^{\alpha_{2}\alpha_{1}}}\   \right],
    \end{aligned}
\end{equation}
in which $\tilde{\mathcal{L}}_{1}^{\alpha_{i}\alpha_{j}}=\langle k_{i}|\otimes\langle k_{i}^{\prime}|^{*}\tilde{\mathcal{L}}_{1}|k_{j}\rangle\otimes|k_{j}^{\prime}\rangle^{*}$.
For $L$-fold degenerate case, the first-order corrections of eigenvalues follow
\begin{equation}\label{}
 \begin{vmatrix}
\tilde{\mathcal{L}}_{1}^{\alpha_{1}\alpha_{1}}-\lambda^{(1)}_{\alpha} & \tilde{\mathcal{L}}_{1}^{\alpha_{1}\alpha_{2}}&......&\tilde{\mathcal{L}}_{1}^{\alpha_{1}\alpha_{L}} \\
\tilde{\mathcal{L}}_{1}^{\alpha_{2}\alpha_{1}} & \tilde{\mathcal{L}}_{1}^{\alpha_{2}\alpha_{2}}-\lambda^{(1)}_{\alpha}&......&\tilde{\mathcal{L}}_{1}^{\alpha_{2}\alpha_{L}} \\
....&....  &......&.... \\
\tilde{\mathcal{L}}_{1}^{\alpha_{L}\alpha_{1}} & \tilde{\mathcal{L}}_{1}^{\alpha_{L}\alpha_{2}}&......&\tilde{\mathcal{L}}_{1}^{\alpha_{L}\alpha_{L}}-\lambda^{(1)}_{\alpha}
 \end{vmatrix}   =0,
\end{equation}
in which $\tilde{\mathcal{L}}^{\alpha_{i}\alpha_{j}}=\langle k_{i}|\otimes\langle k_{i}^{\prime}|^{*}\tilde{\mathcal{L}}_{1}|k_{j}\rangle\otimes|k_{j}^{\prime}\rangle^{*}$. The first-order perturbation results  are shown in Fig. \ref{suppfig1}(a).

\subsubsection{steady state}

As mentioned above, the steady state $\rho_{ss}$ is corresponding to $k=k^{\prime}=-\dfrac{\pi}{2}$. 
The Liouvillian gap $\text{Re}(\lambda_{1})\approx\dfrac{\gamma}{2}(1+\text{sin}(-\dfrac{\pi}{2}+\Delta k))\approx\dfrac{\gamma}{4}(\Delta k)^2=\dfrac{\pi^2\gamma}{L^2}$, aligning with Fig. \ref{fig3}. Specifically, the unique steady state $\rho_{ss}$ takes the form 
\begin{equation}\label{rhoss1}
\begin{aligned}
\rho_{ss}&=|-\dfrac{\pi}{2}\rangle\otimes|-\dfrac{\pi}{2}\rangle^{*} \\
&=\dfrac{1}{L}\sum_{n,m=1}^{L}e^{i\dfrac{\pi}{2}(n-m)}|n\rangle\otimes|m\rangle,
\end{aligned}
\end{equation}
which is highly coherent.
However, steady state $\rho_{ss}$ should maintain periodicity, which demands $e^{i\dfrac{\pi}{2}L}=1$, namely $L=4N$, $N\in \mathcal{Z}$. Another evidence is that  $k=-\dfrac{\pi}{2}$ can be  exactly chosen only when $L=4N, N\in\mathcal{Z}$.

In comparison, for $L\neq 4N$, $N\in \mathcal{Z}$, it is known that $k=\dfrac{2\pi m}{L}$, $m=1,2,...,L$, so $k\neq -\dfrac{\pi}{2}\ (\dfrac{3\pi}{2})$, thus $|-\dfrac{\pi}{2}\rangle\otimes|-\dfrac{\pi}{2}\rangle$ will not be steady state. However, from the perturbation results, it is   justifiable to assume that the steady state takes the form $\rho_{ss}=\sum_{k}\lambda_{k}|k\rangle\otimes|k\rangle^{*}, k=2\pi j/L,j=1,2,...L$. Therefore, the steady state is required to obey $\tilde{\mathcal{L}}\rho_{ss}=(\tilde{\mathcal{L}}_{0}+\gamma\tilde{\mathcal{L}}_{1})\rho_{ss}=0$, which can be simplified as 
\begin{equation}\label{equality}
\begin{aligned}
    &\sum_{k}\dfrac{\lambda_k}{L}(1+\text{sin}k)\sum_{k_1}(1-\text{sin}k_1)|k_1\rangle\otimes|k_1\rangle^{*} \\ 
    &=\sum_{k}\lambda_{k}(1+\text{sin}k)|k\rangle\otimes|k\rangle^{*}.
    \end{aligned}
\end{equation}
According to the Eq. \ref{equality}, it can be readily  inferred that $\lambda_{k}=\dfrac{1}{\sum_{k}\dfrac{1-\text{sin}k}{1+\text{sin}k}}\dfrac{1-\text{sin}k}{1+\text{sin}k}$. 

So, in general, the steady state is $\rho_{ss}=\sum_{k}\lambda_{k}|k\rangle\otimes|k\rangle^{*}$, $\lambda_{k}=\dfrac{1}{\sum_{k}\dfrac{1-\text{sin}k}{1+\text{sin}k}}\dfrac{1-\text{sin}k}{1+\text{sin}k}$. When $L=4N$, all $\lambda_{k}=0$ equal zero except $\lambda_{-\pi/2}=1$, which is consistent with Eq. \ref{rhoss1}.

\subsection{With feedback, OBCs}
\label{appendixsubsection2}

In this section, we also perform the perturbation analysis for the feedback case under OBCs, which elucidates the Lindbladians'  extreme sensitivity of boundary conditions and the localization of eigenmodes under OBCs. For feedback case under OBCs, the Hamiltonian is expressed as   $H=\dfrac{1}{4}\sum_{j=1}^{L-1}(c_{j}^{\dagger}c_{j+1}+c^{\dagger}_{j+1}c_{j})$ and the effective non-Hermitian is given by  $H_{\text{eff}}=H-i\dfrac{\gamma}{2}\sum_{j=1}^{L-1}L_{j}^{\dagger}L_{j}$.
\subsubsection{Liouvillian spectrum}
Likewise, we can perform the same decomposition of the Lindbladian  as Appendix. \ref{appendixsubsection1}, namely
 $\tilde{\mathcal{L}}=\tilde{\mathcal{L}}_{0}+\gamma\tilde{\mathcal{L}}_{1}$, in which $\tilde{\mathcal{L}}_{0}=-i H_{\text{eff}}\otimes I+i I\otimes H_{\text{eff}}^{*}$ and the perturbed part is $\tilde{\mathcal{L}}_{1}=\sum_{j=1}^{L-1}L_j\otimes L_j^*$. The effective non-Hermitian Hamiltonian is $H_{\text{eff}}=H-i\dfrac{\gamma}{2}\sum_{j=1}^{L-1}L^{\dagger}_{j}L_{j}=\sum_{j=1}^{L-1}(\dfrac{1+\gamma}{4}c^{\dagger}_{j}c_{j+1}+\dfrac{1-\gamma}{4}c^{\dagger}_{j+1}c_{j})-i\dfrac{\gamma}{4}(2-n_{L}-n_{1})$. Neglecting the terms $i\dfrac{\gamma}{4}(n_{L}+n_{1})$ of $H_{\text{eff}}$,
 the right eigenvectors of $H_{\text{eff}}$ are approximate to  $|k^{R}\rangle\approx\dfrac{ 1}{\sqrt{L}}\sum_{n=1}^{L}(\sqrt{\dfrac{|1-\gamma|}{1+\gamma}})^ne^{-ikn}|n\rangle$, 
 while the left eigenvectors of  $H_{\text{eff}}$ are  approximately as $\langle k^{L}|\approx\dfrac{1}{\sqrt{L}}\sum_{n=1}^{L}(\sqrt{\dfrac{1+\gamma}{|1-\gamma|}})^ne^{ikn}\langle n|$. The right and left eigenvectors satisfy biorthogonal relation $\langle k^{L}|{k^{\prime}}^{R}\rangle=\delta_{k,k^{\prime}}$. Consequently,  we can obtain $\langle k^{L}|H_{\text{eff}}|k^{R}\rangle\approx\dfrac{1}{2}\sqrt{1-\gamma^2}\ \text{cos}k-i\dfrac{\gamma}{2}$ for $\gamma<1$, while $\langle k^{L}|H_{\text{eff}}|k^{R}\rangle\approx-\dfrac{i}{2}(\sqrt{\gamma^2-1}\  \text{sin}k+\gamma)$ for $\gamma>1$.
 Therefore, the  zeroth-order eigenvalues of the Lindbladian are approximately 
 \begin{equation}\label{OBCFeedbackEigenvalue}
 \begin{aligned}
     \lambda^{(0)}_{k,k^{\prime}}&=\langle k^{L}|\otimes\langle {k^{\prime}}^{L}|^{*}\tilde{\mathcal{L}}_{0}|k^{R}\rangle\otimes |{k^{\prime}}^{R}\rangle^{*} \\ 
    & \approx\left\{\begin{array}{l}
-\dfrac{i}{2}\sqrt{1-\gamma^{2}}(\text{cos}k-\text{cos}k^{\prime})-\gamma, \text{for}\ \gamma<1  \\
-\dfrac{1}{2}\sqrt{\gamma^2-1}(\text{sin}k+\text{sin}k^{\prime})-\gamma, \text{for}\ \gamma>1
\end{array}\right.
\end{aligned}
 \end{equation}

The approximate  zeroth-order eigenvalues of the Liouvillian roughly tells that for $\gamma<1$, the Liouvillian eigenvalues will satisfy $\text{Re}\{\lambda\}=-\gamma, \text{Im}\{\lambda\}\in[-\sqrt{1-\gamma^2},\sqrt{1-\gamma^2}]$, which is roughly fitted with Fig. \ref{fig1}(a), and for $\gamma>1$, the Liouvillian eigenvalues should satisfy $\text{Re}\{\lambda\}=[-\sqrt{{\gamma}^2-1}-\gamma,\sqrt{\gamma^2-1}-\gamma], \text{Im}\{\lambda\}=0$, which is also approximately consistent with Fig. \ref{fig1}(d), where most of  eigenvalues are located at real axis.
For $\gamma=1$, there are huge degeneracy of the Liouvillian eigenvalue $\lambda=1$ in Fig. \ref{fig1}(c), which can be simply understood from Eq. \ref{OBCFeedbackEigenvalue}.

\begin{figure}[h]
\centering
\includegraphics[height=6.0cm,width=8.0cm]{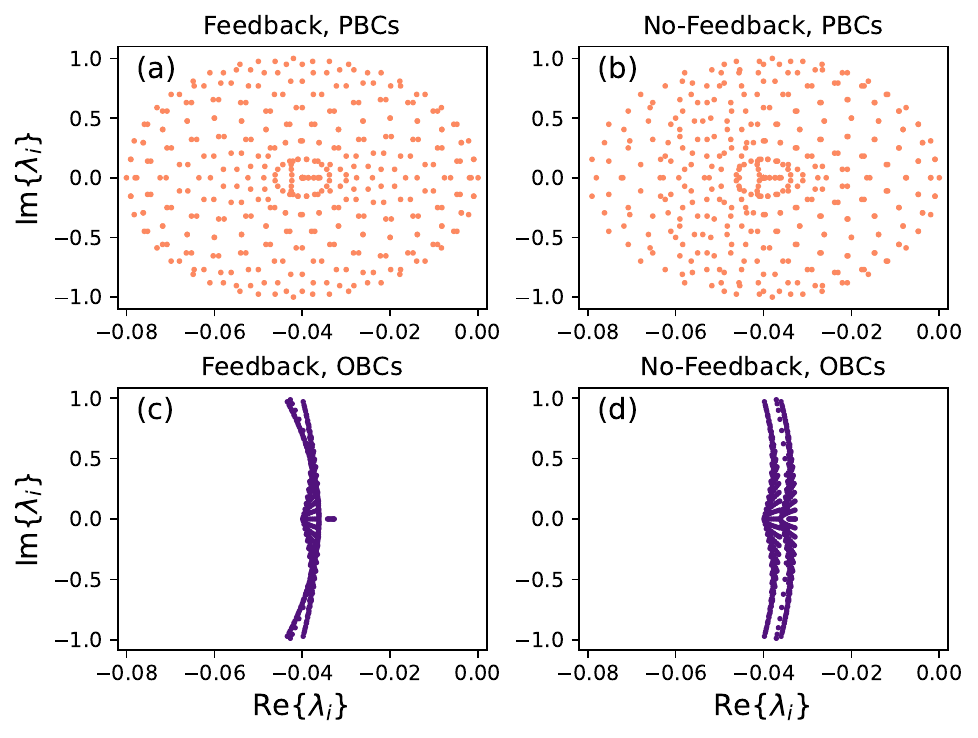}
\caption{The Liouvillian spectrum obtained through first-order degenerate and non-degenerate perturbation analysis. The parameters are $L=20$, $\gamma=0.04$.}
\label{suppfig1}
\end{figure}

Certainly, although being able to capturing some properties of spectrum, the above approximate zeroth-order solution is too inaccurate. Therefore, we also numerically  perform the perturbation analysis without neglecting the terms $i\dfrac{\gamma}{4}(n_{L}+n_{1})$. Interestingly, numerical results indicate that corrections above zeroth-order are also about $O(1/L)$, similar with the case under PBCs. In addition, we find degeneracies are common as well under OBCs. For example, when $\gamma=0.04$, there are  $L^{2}-L$ double degenerate zeroth-order eigenvalues and $L$ non-degenerate zeroth-order eigenvalues. Therefore, we also perform degenerate perturbation theory as mentioned before. The first-order perturbation results  are shown in Fig. \ref{suppfig1}(c).

As for steady state, unlike the PBCs case, we can not  analytically derive out the exact form of the steady state. The numerical results show that the  steady state exhibits  Liouvillian skin effect. Theoretically, because $|k^R\rangle$ is exponentially localized around the left edge,   the resultant approximated zeroth-order eigenvectors $|k^R\rangle\otimes|{k^{\prime}}^R\rangle^{*}$ imply that the eigenmodes under OBCs tend to localize around the corner $|1\rangle\otimes|1\rangle$,  aligning with Fig. \ref{suppfig2}.

\subsection{No feedback, PBCs}
\label{appendixsubsection3}
Compared with the feedback case under PBCs, the Lindblad operators for no-feedback case are modified as   $L_{j}=\dfrac{1}{2}(n_{j}+n_{j+1})+\dfrac{i}{2}(c_{j}^{\dagger}c_{j+1}-c^{\dagger}_{j+1}c_{j})=L^{\dagger}_{j}L_{j}$, while $H$ and $L_{j}^{\dagger}L_{j}$ keep invariant.
Therefore, we can perform the same analysis as Sec. \ref{appendixsubsection1}. 
The zeroth-order eigenvalues are still  $\lambda_{k,k^{\prime}}^{(0)}=-\dfrac{i}{2}(\text{cos}k-\text{cos}k^{\prime})-\dfrac{\gamma}{2}(2+\text{sin}k+\text{sin}k^{\prime})$. Moreover, the first-order corrections are   $\lambda_{k,k^{\prime}}^{(1)}=\dfrac{1}{L}(1+\text{sin}k)(1+\text{sin}k^{\prime})$, indicating that  $\lambda_{k,k^{\prime}}^{(1)}$ is of the order $O(1/L)$.
So the Liouvillian spectrum still approximately satisfies $-2\gamma\leq\text{Re}\{\lambda\}\leq0$ and $-1\leq\text{Im}\{\lambda\}\leq1$. The first-order perturbation results are depicted in Fig. \ref{suppfig1}(b).

Because Lindblad operators $L_{j}$ are Hermitian, it is well known that  $\mathcal{L}\Bbb{I}=0$. In other words, the $\sigma=\dfrac{1}{L}\sum_{j=1}^L|j\rangle\otimes|j\rangle$ satisfies $\tilde{\mathcal{L}}\sigma=0$. Interestingly, we discover that there can exist bistable steady states. Specifically, the same as Sec.  \ref{appendixsubsection1}, we assume the steady state is $\rho_{ss}=\sum_{k}\eta_{k}|k\rangle\otimes|k\rangle^{*}$, then solve $\tilde{\mathcal{L}}\rho_{ss}=0$ and obtain 
\begin{equation}
    \sum_{k}\left[(\dfrac{1}{L}\sum_{k_1}\eta_{k_1}(1+\text{sin}k_1)-\eta_{k})(1+\text{sin}k)|k\rangle\otimes|k\rangle^{*}\right]=0.
\end{equation}
For $L\neq 4N$, it is evident that  $1+\text{sin}k\neq0 \ \forall k$, so it requires  $\eta_{k}=\dfrac{1}{L}\sum_{k_1}\eta_{k_1}(1+\text{sin}k_1)$. Combined with  $\sum_{k}\eta_k=1$, the solutions are $\eta_k=\dfrac{1}{L} \ \forall k$, which tells that  $\rho_{ss}=\Bbb{I}/L$.
While for $L=4N$, it is clear that $\rho_{ss}=\Bbb{I}/L$ is still the solution. However, there is another set of solution. Owing to  $1+\text{sin}(\dfrac{2\pi}{L}\dfrac{3L}{4})=1+\text{sin}(-\dfrac{\pi}{2})=0$, we can set all $\eta_{k}=0$ except for $\eta_{-\pi/2}=1$. Therefore, for $L=4N$, the steady state is 
\begin{equation}
\begin{aligned}
\rho_{ss}&=a_{1}\Bbb{I}/L+a_{2}\dfrac{1}{L}|-\dfrac{\pi}{2}\rangle\otimes|-\dfrac{\pi}{2}\rangle^{*} \\ 
&=a_{1}\dfrac{1}{L}\sum_{n=1}^{L}|n\rangle\otimes|n\rangle+a_{2}\dfrac{1}{L}\sum_{n,m=1}^{L}i^{n-m}|n\rangle\otimes|m\rangle, 
\end{aligned}
\end{equation}
in which $a_{1}+a_{2}=1$ for preserving unity of trace.

\subsection{No feedback, OBCs}
\label{appendixsubsection4}
For no-feedback case under OBCs, $\tilde{\mathcal{L}_{0}}$ is the same as the feedback case under OBCs. So we can perform the same perturbation analysis as Sec. \ref{appendixsubsection2}, which also roughly explains the shape of Liouvillian spectrum. The first-order perturbation results are depicted in Fig. \ref{suppfig1}(d). In addition,  despite the steady state being the trivial maximally mixed state $\Bbb{I}/L$, the majority of eigenmodes still  exhibits localization, which can be roughly understood from the  localization of the approximated zeroth-order eigenvectors.

\begin{figure}[h]
\centering
\includegraphics[height=6.6cm,width=8.8cm]{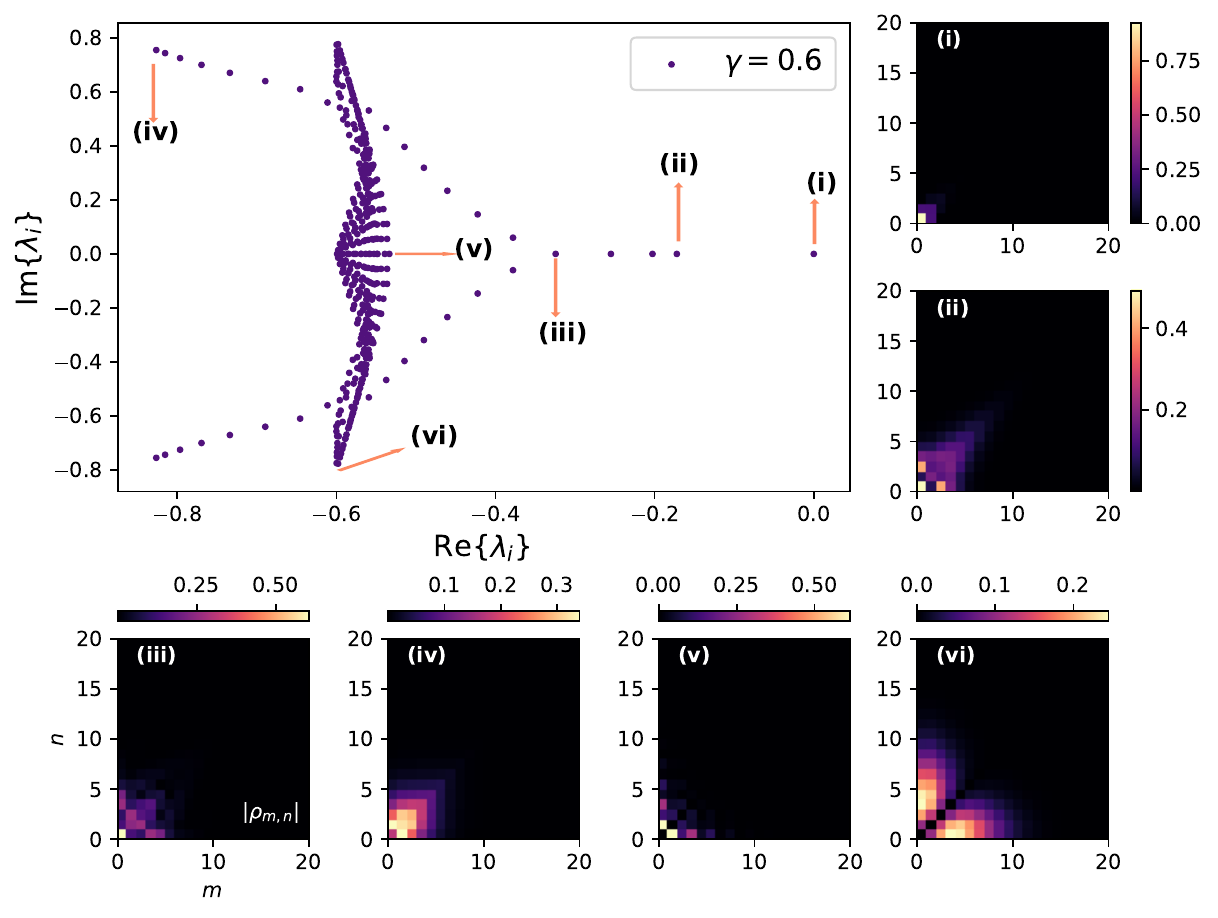}
\caption{Liouvillian spectrum and eigenmodes for feedback case under OBCs. The parameters are $L=20, \gamma=0.6$. Besides steady state, almost all eigenmodes exhibit Liouvillian skin effect. }
\label{suppfig2}
\end{figure}

In summary, by perturbation analysis, it is obvious that the sensitivity of the Lindbladian is inherited from 
the non-Hermitian effective Hamiltonian $H_{\text{eff}}$. Specifically, comparing Fig.  \ref{suppfig1} with Fig. \ref{fig1}(a) and Fig. \ref{fig15}(a), the first-order perturbation results already roughly give correct shape of Liouvillian spectrum. Moreover, the perturbation results also imply that eigenmodes under OBCs will tend to be exponentially localized, which is also consistent with exact diagonalization results in Fig. \ref{fig18} and Fig. \ref{suppfig2}. Of course, there are still some deviations between perturbation and exact diagonalization results, one of which is that steady state can not be acquired by perturbation analysis. 

\section{Additional numerical results}
\label{additionalNumericalResults}

\subsection{Eigenmodes with feedback}

As Fig. \ref{suppfig2} shows, for Lindbladian with feedback under OBCs, besides steady state, almost all  eigenmodes exhibit LSE, in contrast with no-feedback case in Fig. \ref{fig18}.

\subsection{Many-body Liouvillian spectrum}
\begin{figure}[h]
\centering
\includegraphics[height=6.0cm,width=8.0cm]{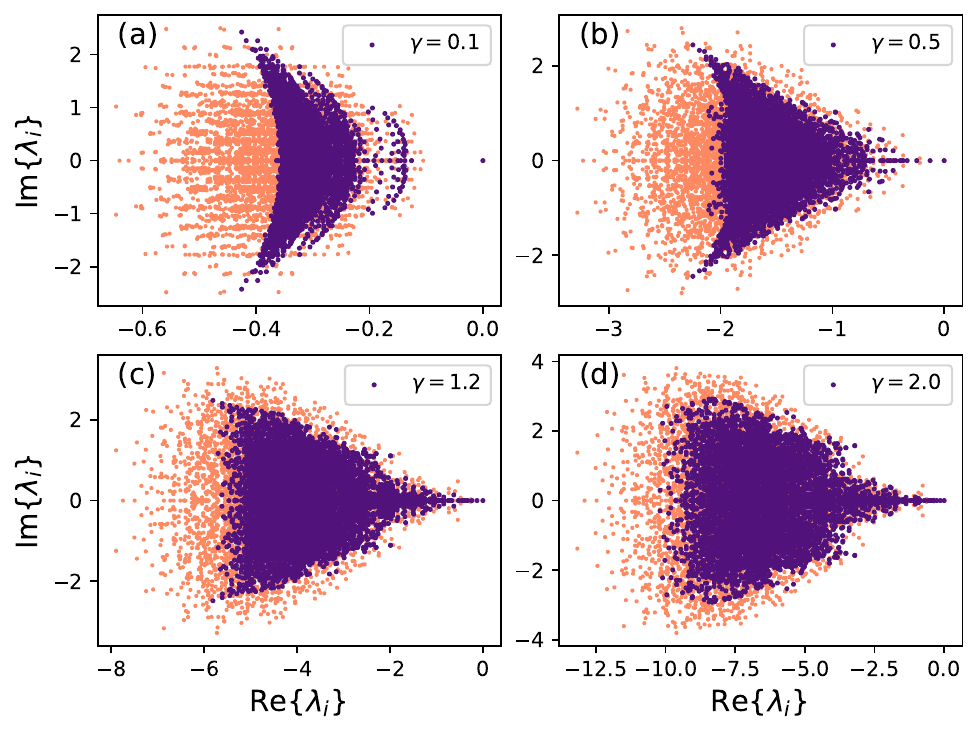}
\caption{Many-body Liouvillian spectrum for feedback case under OBCs (purple) and PBCs (orange). The system size is $L=8$ and particle numbers are $ N=4$.}
\label{suppfig3}
\end{figure}

\begin{figure}[h]
\centering
\includegraphics[height=6.0cm,width=8.0cm]{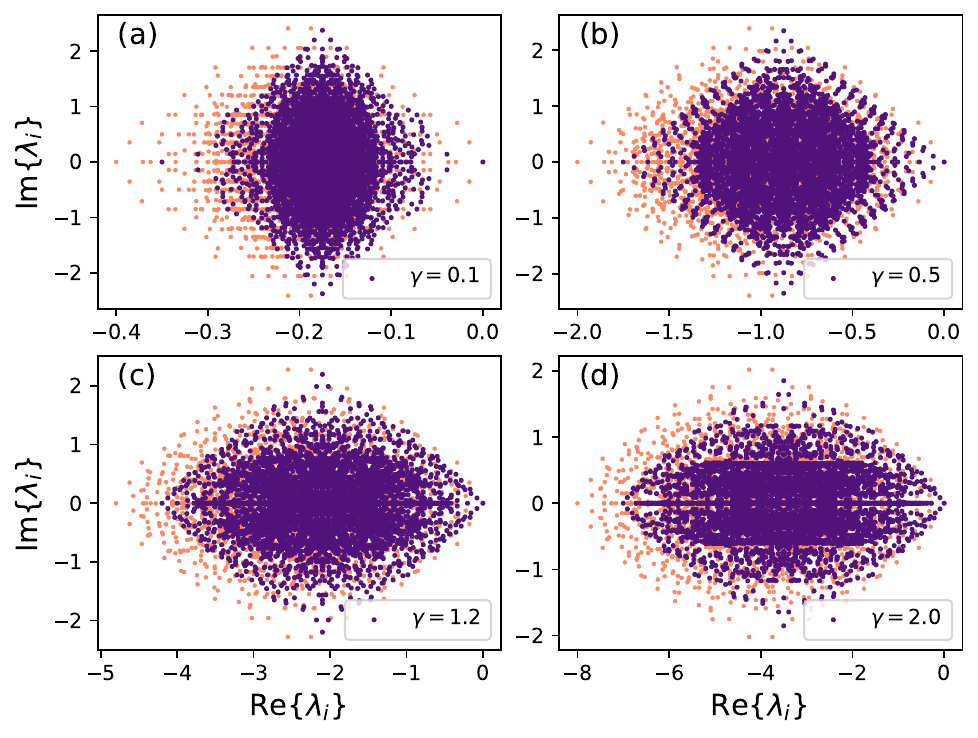}
\caption{Many-body Liouvillian spectrum for no-feedback case under OBCs (purple) and PBCs (orange). The system size is $L=8$ and particle numbers are $ N=4$.}
\label{suppfig4}
\end{figure}

\begin{figure}[h]
\centering
\includegraphics[height=6.0cm,width=8.0cm]{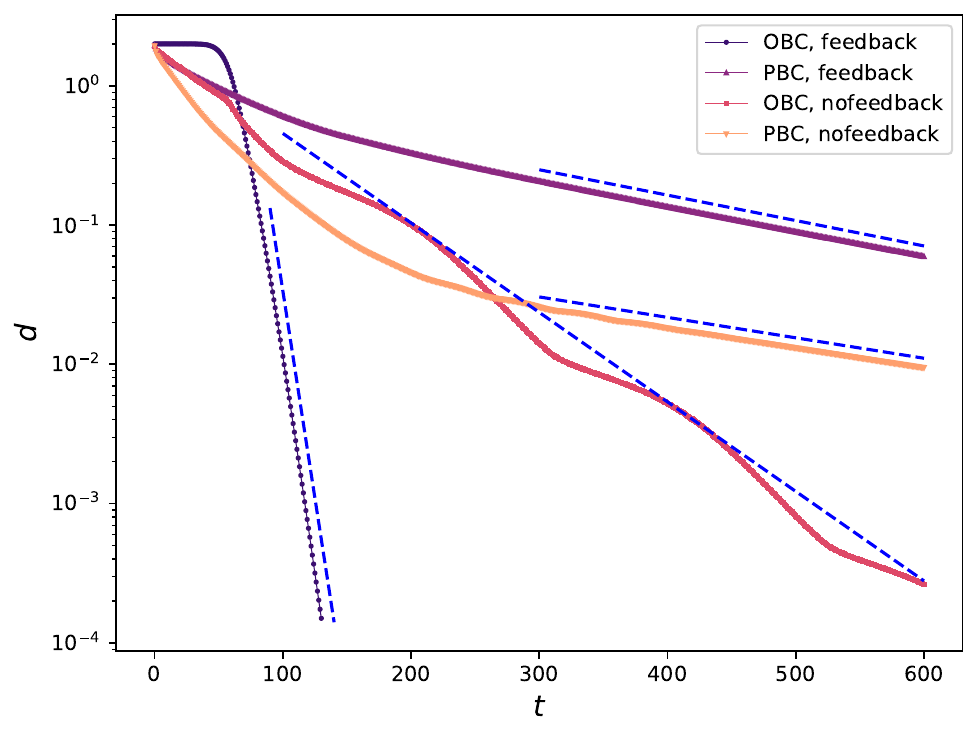}
\caption{Evolution of the distances on a semi-log plot for four cases we studied in this paper. The parameters are $L=30$ and  $\gamma=0.6$. The blue dotted lines represented the asymptotic  regime, in which the distance almost decays as the inverse of the Liouvillian gap. }
\label{suppfig5}
\end{figure}

\begin{figure}[h]
\centering
\includegraphics[height=6.0cm,width=8.0cm]{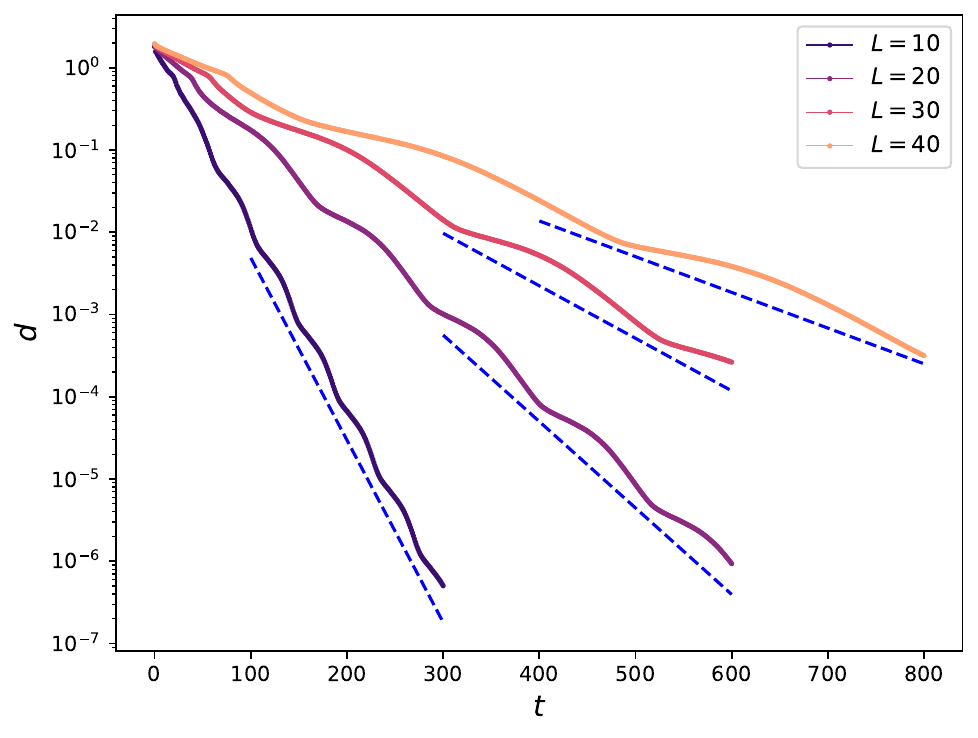}
\caption{Evolution of distances on a semi-log plot for no-feedback case under OBCs. The parameters are $L=10, 20, 30, 40$ and $\gamma=0.6$. }
\label{suppfig6}
\end{figure}

As shown in Fig. \ref{suppfig3} and Fig. \ref{suppfig4}, the PBCs Liouvillian spectra are still distinct from the OBCs Liouvillian spectra for many-body case. Moreover, the many-body Liouvillian spectrum for Lindbladian with feedback  differs considerably  from the Liouvillian spectrum for no-feedback case. Because in the many-body case, the Lindblad operators for feedback case are quartic $L_{i}=\dfrac{1}{2}(n_{i}-n_{i+1})+\dfrac{i}{2}(c^{\dagger}_{i}c_{i+1}+c^{\dagger}_{i+1}c_{i})-n_{i}n_{i+1}$. However,  the Lindblad operators for the no-feedback case are still quadratic $L_{i}=\dfrac{1}{2}(n_{i}+n_{i+1})+\dfrac{i}{2}(c^{\dagger}_{i}c_{i+1}-c^{\dagger}_{i+1}c_{i})$.
Hence, it is worthwhile to investigate the relaxation  behavior in the man-body case and explore possible new phenomena beyond the single-body case. 

Interestingly,  for many-body case, if $L=4N, N\in\mathcal{Z}$, there still  exists bistable steady states for Lindbladian without feedback under PBCs, which can be constructed from bistable steady state in the single-particle case. For example, if $L=4, N=2$, besides $\Bbb{I}/L$, another steady state is $\sum_{k}|-\dfrac{\pi}{2},k\rangle\otimes|-\dfrac{\pi}{2},k\rangle^{*}$ (not normalized).

\subsection{Evolution of distance}

The Fig. \ref{suppfig5} shows typical decay behaviors of distances for both Lindbladians with feedback and without feedback under PBCs and OBCs. The different decay behaviors of distances are evident signatures of boundary-sensitive relaxation behaviors.

As shown in Fig. \ref{suppfig6}, the distances for no-feedback case under OBCs immediately decays as the manner controlled by Liouvillian gap $\Delta$. Therefore, the corresponding relaxation time $\tau$ is given by  $\tau\sim 1/\Delta$.

\end{CJK*}
\end{document}